\documentclass{article}

\usepackage{arxiv}
\usepackage[T1]{fontenc}
\usepackage[utf8]{inputenc}

\usepackage{hyperref}
\usepackage{url}
\usepackage{booktabs}

\usepackage{amsmath,amssymb, amsfonts}
\usepackage{nicefrac}
\usepackage{microtype}
\usepackage{graphicx}
\usepackage{subcaption}
\usepackage[
    backend=biber,
    style=authoryear,
    natbib=true,
]{biblatex}
\usepackage{fancybox}
\cornersize{0.75}

\usepackage{physics}
\usepackage{bm}
\usepackage{siunitx}
\usepackage{todonotes}

\usepackage{minted}

\usepackage{mathtools}
\usepackage[version=4]{mhchem}

\usepackage{todonotes}

\usepackage[verbose]{newunicodechar}

\newcommand{\uwidth}[1]{\makebox[0.5em]{#1}}

\newunicodechar{δ}{\uwidth{\ensuremath{\delta}}}
\newunicodechar{λ}{\uwidth{\ensuremath{\lambda}}}
\newunicodechar{κ}{\uwidth{\ensuremath{\kappa}}}
\newunicodechar{μ}{\uwidth{\ensuremath{\mu}}}
\newunicodechar{φ}{\uwidth{\ensuremath{\varphi}}}
\newunicodechar{ε}{\uwidth{\ensuremath{\epsilon}}}
\newunicodechar{τ}{\uwidth{\ensuremath{\tau}}}

\usepackage{calc}

\newtheorem{theorem}{Theorem}

\newcommand{\trp}{\mathsf{T}}

\newcommand{\ddiv}{\nabla \cdot}

\title{Automated dimensional analysis for PDEs}

\author{%
    Michal HABERA \\
    Department of Engineering\\
    University of Luxembourg\\
    \texttt{michal.habera@uni.lu} \\
    \And
    Andreas ZILIAN \\
    Department of Engineering\\
    University of Luxembourg\\
    \texttt{andreas.zilian@uni.lu} \\
}

\date{June 2026}
\renewcommand{\headeright}{Preprint}
\renewcommand{\undertitle}{Preprint}

\hypersetup{
pdftitle={Automated dimensional analysis for PDEs},
pdfauthor={M.~Habera, A.~Zilian},
}

\begin{document}
\maketitle

\begin{abstract}
  Physical units are fundamental to scientific computing. However, many finite element frameworks
  lack built-in support for dimensional analysis. In this work, we present a systematic framework
  for integrating physical units into the Unified Form Language (UFL).

  We implement a symbolic \texttt{Quantity} class to track units within variational forms. The
  implementation exploits the abelian group structure of physical dimensions. We represent them as
  vectors in $\mathbb{Q}^n$ to simplify operations and improve performance. A graph-based visitor
  pattern traverses the expression tree to automate consistency checks and factorization.

  We demonstrate that this automated nondimensionalization functions as the simplest form of Full
  Operator Preconditioning. It acts as a physics-aware diagonal preconditioner that equilibrates
  linear systems prior to assembly. Numerical experiments with the Navier--Stokes equations show
  that this improves the condition number of the saddle-point matrix. Analysis of Neo-Hooke
  hyperelasticity highlights the detection of floating-point cancellation errors in small
  deformation regimes. Finally, the Poisson--Nernst--Planck system example illustrates the
  handling of coupled multiphysics problems with derived scaling parameters. Although the
  implementation targets the FEniCSx framework, the concepts are general and easily adaptable to
  other finite element libraries using UFL, such as Firedrake or DUNE.
\end{abstract}


\section{Introduction}

Physical units are fundamental to scientific computing, yet many numerical libraries lack built-in
support for dimensional analysis and unit consistency checking. This deficiency is particularly
acute in the field of Finite Element Method (FEM) for the solution of partial differential equations
(PDEs), where complex multiphysics models often involve quantities spanning many orders of magnitude
and diverse physical dimensions.

The importance of robust unit handling in scientific and engineering software cannot be overstated,
as demonstrated by historical failures such as NASA's Mars Climate Orbiter disaster in
1999~\citep{nasa_mco}. The mission's 125 million dollar spacecraft was lost when it entered the
Martian atmosphere at an incorrect angle due to a units mismatch. Ground software produced output in
pound-force-seconds, while the spacecraft navigation system expected newton-seconds. This costly
error highlights why strong type checking and dimensional analysis should be fundamental aspects of
scientific computing frameworks. For more examples of failures and accidents caused by mistakes in
physical units, see \citet{p1935r0}.

Our framework addresses this gap by providing a systematic approach to handle units in modeling
based on the Unified Form Language (UFL). It enables automatic nondimensionalization, dimensional
consistency verification, and facilitates the application of the Buckingham Pi theorem for
identifying dimensionless groups in physical systems directly from the variational formulation.

\subsection{Existing libraries for physical units}

Several software packages have attempted to address the problem of unit handling in scientific
computing. Pint~\citep{pint} is a Python package for handling physical quantities with units,
providing arithmetic operations and unit conversions. The Boost.Units C++ template library~\citep{boostunits}
implements compile-time dimensional analysis. For Julia, there are Unitful.jl~\citep{unitful} and
DynamicQuantities.jl~\citep{dynamicquantities}. Julia is a very promising language for unit analysis
given its just-in-time (JIT) compilation and dynamic type deduction. For Python alone, other
packages worth mentioning are Astropy.units~\citep{astropy}, Natu~\citep{Davies2025Natu},
Buckingham~\citep{Mdipierro2025Buckingham}, Magnitude~\citep{Reyero2025Magnitude}, and
Python-quantities~\citep{Dale2025Python}.

SymPy~\citep{sympy} provides a comprehensive units module that combines symbolic mathematics with
dimensional analysis capabilities, allowing for algebraic manipulation of expressions containing
physical quantities with units. Despite this capability, it is primarily designed for symbolic
mathematics rather than numerical computation or finite element assembly.

Recent efforts in the C++ community include the comprehensive proposal P1935R0~\citep{p1935r0}, ``A
C++ Approach to Physical Units'', which aims to standardize physical units support for C++. This
proposal introduces a compile-time dimensional analysis system with quantity types, powerful unit
conversions, and user-defined units. It enables operations such as:

\begin{minted}{cpp}
static_assert(10km / 2 == 5km);     // numeric operations
static_assert(1h == 3600s);         // unit conversions
static_assert(1km / 1s == 1000mps); // dimension conversions
\end{minted}

Additional tools are reviewed in \citet[\S 1.1]{Langtangen2016Scaling}, which provides
a comprehensive treatment of scaling and nondimensionalization techniques for differential equations.

However, the above-mentioned packages primarily focus on standalone calculations rather than integration
with PDE solvers. In this work, we focus on PDEs and systems of PDEs as the governing equations for multiphysics
problems.

%%%%%%%%%%%%%%%%%%%%%%%%%%%%%%%%%%%%%%%%%%%%%%%%%%%%%%%%%%%%%%%%%%%%%%%%%%%%%%%
%%%%%%%%%%%%%%%%%%%%%%%%%%%%%%%%%%%%%%%%%%%%%%%%%%%%%%%%%%%%%%%%%%%%%%%%%%%%%%%

\section{Brief mathematical background}

\subsection{Definitions}

\emph{Quantity} (or \emph{physical quantity}) is a property of a physical system that can be numerically measured (directly or indirectly).
Examples of physical quantities are width, weight, event duration, viscosity, velocity, etc.

\emph{Dimension} (or \emph{physical dimension}) of a quantity describes what kind of quantity it is. It expresses our abstract understanding
of the nature of the quantity. In fact, the number of base physical dimensions depends on what type of quantities we want to describe and what
we know about them. Examples of physical dimensions are length, mass, time, electric current, etc.

\emph{Unit} (or \emph{physical unit}) is a specific, agreed-upon scale used to measure a physical quantity. Examples of
physical units are meter, kilogram, second, ampere, pound, hour, foot, etc.

\subsection{Dimensional analysis principles}
Dimensions of physical quantities can be expressed as products of powers of base dimensions.
For a quantity $q$, we can write
\begin{equation}
  [q] = \text{N}^a \text{I}^b \text{L}^c \text{J}^d \text{M}^e \Theta^f \text{T}^g,
\end{equation}
where N, I, L, J, M, $\Theta$, and T represent the seven base dimensions of the International System of Units (SI)
for amount of substance, electric current, length, luminous intensity, mass, temperature, and time, respectively,
see \citet{Iso80000}. The operator $[\cdot]$ denotes the dimension of the respective quantity.
\begin{table}[h]
  \centering
  \caption{Overview of SI units showing the seven base units and common derived units with their dimension vectors.}
  \label{tab:si_units}
  \begin{tabular}{@{}llcccccccc@{}}
    \toprule
    \textbf{Category}   & \textbf{Physical quantity} & \textbf{Unit}  &
    \multicolumn{7}{c}{\textbf{Dimension vector}}                                                                                                                   \\
    \cmidrule(lr){4-10} &                            &                & \textbf{N} & \textbf{I} & \textbf{L} & \textbf{J} & \textbf{M} & $\bm{\Theta}$ & \textbf{T} \\
    \midrule
    Base units          & Amount of substance        & mole (mol)     & 1          & 0          & 0          & 0          & 0          & 0             & 0          \\
                        & Electric current           & ampere (A)     & 0          & 1          & 0          & 0          & 0          & 0             & 0          \\
                        & Length                     & meter (m)      & 0          & 0          & 1          & 0          & 0          & 0             & 0          \\
                        & Luminous intensity         & candela (cd)   & 0          & 0          & 0          & 1          & 0          & 0             & 0          \\
                        & Mass                       & kilogram (kg)  & 0          & 0          & 0          & 0          & 1          & 0             & 0          \\
                        & Temperature                & kelvin (K)     & 0          & 0          & 0          & 0          & 0          & 1             & 0          \\
                        & Time                       & second (s)     & 0          & 0          & 0          & 0          & 0          & 0             & 1          \\
    \midrule
    Derived units       & Velocity                   & m/s            & 0          & 0          & 1          & 0          & 0          & 0             & -1         \\
                        & Acceleration               & m/s$^2$        & 0          & 0          & 1          & 0          & 0          & 0             & -2         \\
                        & Force                      & newton (N)     & 0          & 0          & 1          & 0          & 1          & 0             & -2         \\
                        & Energy/Work                & joule (J)      & 0          & 0          & 2          & 0          & 1          & 0             & -2         \\
                        & Power                      & watt (W)       & 0          & 0          & 2          & 0          & 1          & 0             & -3         \\
                        & Pressure                   & pascal (Pa)    & 0          & 0          & -1         & 0          & 1          & 0             & -2         \\
                        & Frequency                  & hertz (Hz)     & 0          & 0          & 0          & 0          & 0          & 0             & -1         \\
                        & Electric charge            & coulomb (C)    & 0          & 1          & 0          & 0          & 0          & 0             & 1          \\
                        & Voltage                    & volt (V)       & 0          & -1         & 2          & 0          & 1          & 0             & -3         \\
                        & Resistance                 & ohm ($\Omega$) & 0          & -2         & 2          & 0          & 1          & 0             & -3         \\
    \bottomrule
  \end{tabular}
\end{table}

It is crucial to distinguish between dimensions and units when discussing dimensional analysis. A
dimension represents a physical quantity's nature (e.g., length, time, mass), while a unit is a
specific scale used to measure that dimension (e.g., meters, hours, tons). Two quantities
can have the same dimension but different units.

Mathematically, the set of dimensions forms a free abelian (commutative) group under multiplication,
which can be represented as a vector space over the field of rational numbers.\footnote{The mathematical
  structure can be formalized further, see \citet{Tao2012Dimensional}, but for the purposes of
  practical implementation we focus on the vector space representation.}
Each physical
dimension can be expressed in a chosen base system as a vector in $\mathbb Q^d$.
For the SI system, we have $d=7$ base dimensions, so the dimension of a quantity $q$ can be represented as some
$\mathbf d(q) = (a, b, c, d, e, f, g)^\trp \in \mathbb{Q}^7$, where the components represent the exponents
of the seven SI base dimensions mentioned earlier. For instance
\begin{equation}
  \begin{aligned}
    [\text{velocity}] & = \text{L}\text{T}^{-1},           & \mathbf d(\text{velocity}) & = (0, 0, 1, 0, 0, 0, -1)^\trp, \\
    [\text{force}]    & = \text{M}\text{L}\text{T}^{-2},   & \mathbf d(\text{force})    & = (0, 0, 1, 0, 1, 0, -2)^\trp, \\
    [\text{energy}]   & = \text{M}\text{L}^2\text{T}^{-2}, & \mathbf{d}(\text{energy})  & = (0, 0, 2, 0, 1, 0, -2)^\trp.
  \end{aligned}
\end{equation}

Operations on physical quantities correspond to operations in this vector space:
\begin{itemize}
  \item multiplication of quantities: $[q_1 \cdot q_2] = [q_1] \cdot [q_2]$ corresponds to dimensional vector
        addition
        $$\mathbf d(q_1 \cdot q_2) = \mathbf d(q_1) + \mathbf d(q_2),$$
  \item division of quantities: $[q_1 / q_2] = [q_1] / [q_2]$ corresponds to dimensional vector subtraction
        $$\mathbf d(q_1 / q_2) = \mathbf d(q_1) - \mathbf d(q_2),$$
  \item powers of quantities: $[q^n] = [q]^n$ corresponds to dimensional scalar multiplication
        $$\mathbf d(q^n) = n \cdot \mathbf d(q)$$
        for $n \in \mathbb Q$.
\end{itemize}

Dimensionless quantities (pure numbers) correspond to the identity element in this vector space, which is a vector
with all components equal to zero:
\begin{equation}
  \mathbf d(\text{dimensionless}) = (0, 0, 0, 0, 0, 0, 0)^\trp.
\end{equation}

Two physical quantities can be added or compared only if they have the same dimension vector.
This is the mathematical foundation of the requirement for dimensional homogeneity in physical equations.

\subsection{Buckingham Pi Theorem}

The Buckingham Pi theorem~\citep{buckingham1914physically} provides a rigorous mathematical
foundation for dimensional analysis. Consider a physical system described by $n$ physical quantities
$\{q_1, q_2, \ldots, q_n\}$ involving $k$ physical dimensions. Let $\mathbf{M} \in \mathbb{Q}^{k \times n}$
be the dimensional matrix, where each element $M_{ij}$ represents the exponent of the $i$-th base
dimension in quantity $q_j$. The dimension of quantity $q_j$ can be expressed as
\begin{equation}
  [q_j] = \prod_{i=1}^{k} [D_i]^{M_{ij}}
\end{equation}
where $[D_i]$ represents the $i$-th base dimension. Or, equivalently, the $j$-th column of $\mathbf{M}$
corresponds to the dimension vector $\mathbf d(q_j)$ of quantity $q_j$:
\begin{equation}
  \mathbf M =
  \begin{pmatrix}
    |              & |              &        & |              \\
    \mathbf d(q_1) & \mathbf d(q_2) & \ldots & \mathbf d(q_n) \\
    |              & |              &        & |
  \end{pmatrix}.
\end{equation}

\begin{theorem}[Buckingham Pi]
  Let a physical system be described by a relation $F(q_1, q_2, \ldots, q_n) = 0$ involving $n$ physical quantities
  that depend on $k$ base dimensions. Then the system can be equivalently described by some $G(\Pi_1, \Pi_2, \ldots, \Pi_p) = 0$,
  where $\ker(\mathbf M) = \text{span}\{\bm\alpha_1, \bm\alpha_2, \ldots, \bm\alpha_p\}$ and each $\bm\alpha_j$ is the
  exponent vector of the dimensionless group $\Pi_j = \prod_{i=1}^n q_i^{\alpha_{ji}}$. The $p = \dim(\ker(\mathbf M))$ independent dimensionless
  quantities are called dimensionless groups, and their exponent vectors form a basis for the kernel of the dimensional matrix $\mathbf M$.
\end{theorem}

Equivalently, by the rank--nullity theorem, $p = \dim(\ker(\mathbf M)) = n - \rank(\mathbf{M})$.
The dimensionless groups $\{\Pi_1, \Pi_2, \ldots, \Pi_p\}$, also called $\Pi$-groups, correspond to the kernel of the
matrix $\mathbf{M}$. If $\{\bm{\alpha}_1, \bm{\alpha}_2, \ldots, \bm{\alpha}_p\}$ forms a basis
for $\ker(\mathbf{M})$, then each dimensionless group $\Pi_j$ can be expressed as
\begin{equation}
  \Pi_j = \prod_{i=1}^{n} q_i^{\alpha_{ji}}
\end{equation}
where $\alpha_{ji}$ is the $i$-th component of vector $\bm{\alpha}_j$.

We need to stress at this point that dimensionless groups are not unique. Any function of
dimensionless groups is also dimensionless. Therefore, the choice of basis for $\ker(\mathbf{M})$
is not unique and different choices lead to different sets of dimensionless groups. However,
all choices are equivalent in the sense that they span the same space of dimensionless quantities.
The non-uniqueness of dimensionless groups could be eliminated by additional constraints. This is explored in
later sections where we discuss nondimensionalization of specific PDEs.

\paragraph{Example (Flow around a sphere).}

We include a simple example that illustrates the application of the Buckingham Pi theorem.
Consider the drag force $F_D$ on a sphere of diameter $D$ moving through a fluid with velocity $v$,
density $\rho$, and dynamic viscosity $\mu$. The relevant quantities are $\{F_D, D, v, \rho, \mu\}$
with dimensions
\begin{equation}
  \begin{aligned}
    [F_D]  & = \text{L} \text{M} \text{T}^{-2}, \quad [D] = \text{L}, \quad [v] = \text{L} \text{T}^{-1},                                         \\
    [\rho] & = \text{L}^{-3} \text{M}, \quad [\mu]                                                        = \text{L}^{-1} \text{M} \text{T}^{-1}.
  \end{aligned}
\end{equation}

The dimensional matrix for this system is
\begin{equation}
  \mathbf{M} =
  \begin{pmatrix}
    0  & 0 & 0  & 0  & 0  \\
    0  & 0 & 0  & 0  & 0  \\
    1  & 1 & 1  & -3 & -1 \\
    0  & 0 & 0  & 0  & 0  \\
    1  & 0 & 0  & 1  & 1  \\
    0  & 0 & 0  & 0  & 0  \\
    -2 & 0 & -1 & 0  & -1
  \end{pmatrix}
\end{equation}
where the rows correspond to the dimensions N, I, L, J, M, $\Theta$ and T, and the columns correspond to
the quantities $F_D$, $D$, $v$, $\rho$, and $\mu$, respectively.

Since $\rank(\mathbf{M}) = 3$ and there are $n=5$ quantities, the Buckingham Pi theorem establishes
$p = n - \rank(\mathbf{M}) = 2$ dimensionless groups. Computing a basis for the nullspace of
$\mathbf{M}$ yields
\begin{equation}
  \ker(\mathbf{M}) = \text{span}\left\{
  \begin{pmatrix}
    1, -2, -2, -1, 0
  \end{pmatrix}^\trp,
  \begin{pmatrix}
    0, 1, 1, 1, -1
  \end{pmatrix}^\trp
  \right\},
\end{equation}
which corresponds to the dimensionless groups
\begin{equation}
  \begin{aligned}
    \Pi_1 & = \frac{F_D}{\rho v^2 D^2} \quad \text{(drag coefficient, $C_D$)}, \\
    \Pi_2 & = \frac{\rho v D}{\mu} \quad \text{(Reynolds number, Re)}.
  \end{aligned}
\end{equation}

This formulation has profound implications for modeling physical systems: it reduces the number of
parameters needed to describe the system and reveals the fundamental scaling relationships between
variables. Any physically meaningful equation must be expressible in terms of these dimensionless
groups, providing a powerful indication of insightful forms of governing equations and facilitating both
experimental design and numerical simulation.

%%%%%%%%%%%%%%%%%%%%%%%%%%%%%%%%%%%%%%%%%%%%%%%%%%%%%%%%%%%%%%%%%%%%%%%%%%%%%%%
%%%%%%%%%%%%%%%%%%%%%%%%%%%%%%%%%%%%%%%%%%%%%%%%%%%%%%%%%%%%%%%%%%%%%%%%%%%%%%%

\section{Implementation in a domain specific language for PDEs}

In this section, we present the main theoretical concepts and their implementation for
handling physical units and dimensional analysis within PDE contexts. We focus on the Unified Form Language (UFL)~\citep{ufl},
which is widely used for defining variational forms in finite element frameworks, see the next section for a brief overview.

The implementation is part of the Dolfiny package~\citep{Zilian2021Dolfiny}, which provides a set of extensions and utilities
for the FEniCSx \citep{fenics} finite element framework. However, it is easy to adapt the concepts to other
finite element frameworks that use UFL as their form language, such as Firedrake~\citep{firedrake} or DUNE~\citep{Sander2020Dune}.

The framework introduces a \texttt{Quantity} class to represent physical quantities with associated units. The core of the
framework is based on three main components:
\begin{enumerate}
  \item \emph{Transformation}. A visitor pattern that traverses UFL expression trees to inject unit information
        and reference scales into the expressions.
  \item \emph{Factorization}. A mechanism to extract and manipulate unit factors from expressions, enabling
        dimensional consistency checks and simplifications.
  \item \emph{Normalization}. A procedure to derive dimensionless forms of variational forms by selecting a
        reference factor and dividing through the entire expression.
\end{enumerate}

In the following subsections, we explain the Unified Form Language, the \texttt{Quantity} class,
and the three main components of our framework in more detail.

\subsection{The Unified Form Language}

The Unified Form Language (UFL)~\citep{ufl} is a domain-specific language for declaring finite
element variational forms and functionals in a notation resembling mathematical syntax. It serves as
the front-end for finite element frameworks like FEniCSx~\citep{fenics},
Firedrake~\citep{firedrake} and DUNE~\citep{Sander2020Dune}, enabling users to express PDEs in a form that closely mimics their
mathematical representation while remaining computer-interpretable.

UFL represents mathematical expressions as Directed Acyclic Graphs (DAG), where each node
corresponds to an operation or a terminal object. These expressions can include spatial differential operators
(gradient, divergence, curl, etc.), geometric quantities (normal vectors, mesh coordinates, Jacobian matrix of a transformation, etc.),
and algebraic operations (addition, multiplication, division, etc.).

\paragraph{Example (Poisson equation in UFL).}

To represent the weak form of the Poisson equation
\begin{equation}
  -\Delta u = f,
\end{equation}
one would write the abstract variational formulation: find $u \in U$ such that
\begin{equation}
  \begin{aligned}
    a(u, \delta u) & = L(\delta u)                                         \quad \forall \delta u \in U,                       \\
    a(u, \delta u) & = \int_{\Omega} \nabla u \cdot \nabla \delta u \, dx, \quad L(\delta u) = \int_{\Omega} f \delta u \, dx,
  \end{aligned}
  \label{eq:poisson}
\end{equation}
where $U$ is an appropriate function space, $\delta u \in U$ is a test function, and $\Omega \subset \mathbb R^d$ is the problem domain.
In UFL, this translates to the code shown in Fig.~\ref{fig:poisson-ufl}.

\begin{figure}[h]
  \begin{minted}[fontsize=\footnotesize, linenos, frame=single]{python}
u =  ufl.TrialFunction(U)
δu = ufl.TestFunction(U)
a =  ufl.inner(ufl.grad(u), ufl.grad(δu)) * ufl.dx
L =  ufl.inner(f, δu) * ufl.dx
\end{minted}
  \caption{Implementation of the Poisson equation weak form in the Unified Form Language (UFL).}
  \label{fig:poisson-ufl}
\end{figure}

UFL provides a rich type system for representing different mathematical objects, including
functions, vectors, tensors, and differential operators. It also includes a symbolic differentiation
system that can automatically compute symbolic and functional derivatives, which is essential for implementing
nonlinear solvers and adjoint-based optimization.

Its expression tree structure is particularly suitable for our unit transformation approach, as it
allows systematic traversal and manipulation of mathematical expressions. We leverage UFL's
\texttt{MultiFunction} visitor pattern to implement dimensional analysis and unit transformations
while preserving the mathematical structure of expressions.

%%%%%%%%%%%%%%%%%%%%%%%%%%%%%%%%%%%%%%%%%%%%%%%%%%%%%%%%%%%%%%%%%%%%%%%%%%%%%%%

\subsection{Quantity class}

We define a \texttt{Quantity} class that extends \texttt{ufl.Constant} with unit information.
The \texttt{ufl.Constant} represents a value that is constant over the domain.
Each \texttt{Quantity} stores a symbolic representation (based on \texttt{sympy.Symbol}) alongside its physical
unit using SymPy's physical units module. These quantities serve as reference scales for dimensional
analysis and unit tracking.

The advantage of using SymPy's unit system is that it already implements a comprehensive set of physical units,
including SI base units, derived units, and support for unit conversions. This allows us to leverage
SymPy's capabilities for unit arithmetic and conversions within our framework.

\paragraph{Example (Quantities with units for the Poisson equation).}

The following example demonstrates an implementation of the Poisson equation \eqref{eq:poisson} when used
as a model for electric potential distribution $u$. A reference electric potential, $u_\text{ref}$, which has
dimensions $[u_\text{ref}] = \text{I}^{-1} \text{L}^2 \text{M}^1 \text{T}^{-3}$ takes a specific value
$u_\text{ref} = \SI{1}{\volt}$ when expressed based on the derived unit volt, as shown in Fig.~\ref{fig:quantity-definition}.
On the right-hand side of the equation, we have a source term $f$ with dimensions
$[f] = \text{I}^{-1} \text{M}^1 \text{T}^{-3}$, represented by the reference quantity
$f_\text{ref} = \SI{1}{\volt\per\meter\squared}$. The characteristic length scale of the domain is represented by
$l_\text{ref} = \SI{1}{\meter}$ with dimensions $[l_\text{ref}] = \text{L}$.

\begin{figure}[h]
  \begin{minted}[fontsize=\footnotesize, linenos, frame=single]{python}
import sympy.physics.units as syu
from sympy import Symbol
from dolfiny.units import Quantity
# ...
u_ref = Quantity(mesh, 1.0, syu.volt, "u_ref")
f_ref = Quantity(mesh, 1.0, syu.volt * syu.meter ** (-2), "f_ref")
l_ref = Quantity(mesh, 1.0, syu.meter, "l_ref")
\end{minted}
  \caption{Example of defining reference quantities with physical units using the \texttt{Quantity} class.}
  \label{fig:quantity-definition}
\end{figure}

As a bonus, SymPy's unit system allows the definition of quantities using derived units and metric
prefixes, so we can equivalently define $u_\text{ref} = \SI{1000}{\milli\volt}$, etc.

%%%%%%%%%%%%%%%%%%%%%%%%%%%%%%%%%%%%%%%%%%%%%%%%%%%%%%%%%%%%%%%%%%%%%%%%%%%%%%%

\subsection{Transformation}
\label{sec:transformation}

All UFL expressions are dimensionless: their DAG nodes carry no unit information, so unit
consistency checks and dimensional analysis are not possible. To introduce physical units and
dimensions, we traverse the DAG with a \texttt{UnitTransformer} that replaces selected nodes by
scaled expressions including \texttt{Quantity} objects. Each symbol or operator follows a
transformation rule. Some operators (e.g. the spatial gradient $\nabla$) have built-in rules
depending on the characteristic length of the mesh, $l_{\mathrm{ref}}$. The transformation step injects explicit scaling
factors into the expression DAG to account for the reference quantities.

A special rule is required to transform integrals. For an integral over a domain
$\Omega \subset \mathbb R^d$ or a boundary $\Gamma \subset \mathbb R^{d-1}$ we apply
\begin{equation}
  \begin{aligned}
    \int_\Omega \, \mathrm dx & \mapsto l_\mathrm{ref}^d \int_\Omega \, \mathrm dx, \quad
    \int_\Gamma \, \mathrm ds & \mapsto l_\mathrm{ref}^{d-1} \int_\Gamma \, \mathrm ds
  \end{aligned}
\end{equation}
where $d$ is the topological dimension of the domain. This scaling accounts for the change of units
in the measure of integration.

It is important to note that UFL does not have a built-in support for time derivatives, so time derivative
operators are not handled in the current implementation of the \texttt{UnitTransformer}. The user is expected
to discretize the time derivatives manually, so the insertion of the reference time scale must be
done by the user as well. Integration with automated time discretization frameworks built on top of UFL,
such as Irksome \citep{Farrell2020Irksome}, is a topic for future work.

\paragraph{Example (Transforming the Poisson equation).}

For the Poisson example, the transformation reads
\begin{equation}
  \begin{aligned}
    u     \mapsto u_{\mathrm{ref}} u, \quad
    f     \mapsto f_{\mathrm{ref}} f, \quad
    \delta u     \mapsto u_{\mathrm{ref}} \delta u, \quad
    \nabla \mapsto \frac{1}{l_{\mathrm{ref}}} \nabla.
  \end{aligned}
\end{equation}

Here, $u_{\mathrm{ref}}, f_{\mathrm{ref}}$, and $l_{\mathrm{ref}}$ are \texttt{Quantity} objects
carrying the reference units for the electric potential, source term, and length scale,
respectively.

\begin{figure}[h]
  \centering
  \begin{subfigure}[b]{0.45\textwidth}
    \centering
    \includegraphics{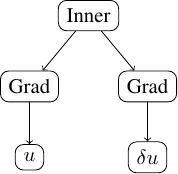}
    \caption{Original DAG: no unit factors present.}
    \label{fig:original-dag}
  \end{subfigure}
  \begin{subfigure}[b]{0.45\textwidth}
    \centering
    \includegraphics{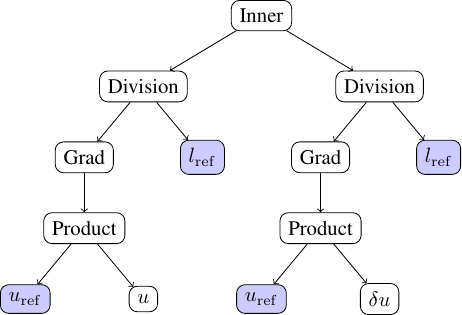}
    \caption{Transformed DAG: each gradient is rescaled by $l_{\mathrm{ref}}^{-1}$
      and terminal nodes with $u_{\mathrm{ref}}$.}
    \label{fig:transformed-dag}
  \end{subfigure}
  \caption{Comparison of the UFL expression DAG before and after inserting quantity scaling factors.
    Blue nodes show the added \texttt{Quantity} objects and their effect on the computation graph.}
  \label{fig:dag-comparison}
\end{figure}

Figure \ref{fig:dag-comparison} illustrates this process for the integrand of the bilinear form in
the Poisson problem, $\nabla u \cdot \nabla \delta u$. In the original DAG (Fig.~\ref{fig:original-dag}),
the two gradient nodes \ovalbox{Grad} feed directly into the inner product node \ovalbox{Inner}. In
the transformed DAG (Fig.~\ref{fig:transformed-dag}), each gradient is first divided by the length
reference $l_\text{ref}$ (to nondimensionalize the derivative) and then multiplied by the field
reference $u_\text{ref}$. The blue-shaded nodes denote the inserted
\texttt{Quantity} factors, while the white nodes show the original UFL operators. Nodes labeled
\ovalbox{Division} and \ovalbox{Product} were introduced by the transformer.

%%%%%%%%%%%%%%%%%%%%%%%%%%%%%%%%%%%%%%%%%%%%%%%%%%%%%%%%%%%%%%%%%%%%%%%%%%%%%%%

\subsection{Factorization}
\label{sec:factorization}

The main idea in checking dimensional consistency and dimensional analysis lies in the
\texttt{Quantity} factorization step. In this step, a DAG traversal class called
\texttt{QuantityFactorizer} applies homogeneous factorization of selected UFL expressions.

Recall that a function $f: V \rightarrow W$ is called positively homogeneous of degree $k \in \mathbb R$, if
\begin{equation}
  f(\alpha v) = \alpha^k f(v)
\end{equation}
holds for all $\alpha \in \mathbb R, \alpha > 0$ and all $v \in V$. Homogeneous factorization is a
transformation that factors the $\alpha^k$ from the DAG and explicitly represents the factor
$\alpha^k$ for each selected UFL expression.

The factorization is executed in a visitor pattern that traverses the DAG in a
post-order manner (child nodes before parents). Internally, the factors for each DAG node store
a vector $\mathbf d \in \mathbb Q^n$, where $n$ is the number of \texttt{Quantity} objects provided
to the factorizer. The vector represents homogeneous dependence of the respective node with respect
to the list of \texttt{Quantity} objects provided to the factorizer.

Special handling and checks are required for the sum operator. Consider a weighted sum
\begin{equation}
  f(v) = \alpha\,g(v) + \beta\,h(v)
\end{equation}
which is nonhomogeneous in $v$. We define two types of inconsistencies that can arise during the
factorization of the sum:
\begin{enumerate}
  \item \emph{Inconsistent factors}: $\alpha \neq \beta$, and, as a special case,
  \item \emph{Inconsistent dimensions}: $[\alpha]\neq[\beta]$.
\end{enumerate}
When either of these inconsistencies are detected, the factorization process raises an exception. The two
types of inconsistencies have different implications that need to be treated differently.

\emph{Inconsistent factors} indicate a mismatch in the scales and magnitudes of the summands. This could have
implications for the numerical conditioning or numerical stability of the problem, and ideally a transformation that
equilibrates all scales in the sum is required. In principle, the following strategies exist to handle inconsistent factors:
\begin{enumerate}
  \item Choice of a common reference scale, such that $\alpha = \beta$. This involves an assumption that both factors are
        of similar magnitude and can be rescaled to a common reference. Typically, there is enough freedom in the
        equations to choose the reference length, time, or field scale to achieve this.
        A common example of this approach is the choice of reference time scale computed from the reference length
        and velocity scale in convection dominated problems.
  \item Exploit the linearity of the equation and treat each term $\alpha g(v)$ and $\beta h(v)$ separately. While this does
        not always resolve the possible numerical and conditioning issues, it explicitly acknowledges the presence of different scales in the problem.
        It suggests that some terms may be neglected in the overall formulation if their contribution is small compared to others.
        This approach is exercised in the examples on the Neo-Hooke elasticity and Poisson--Nernst--Planck equations in later sections.
  \item Introduce auxiliary variables to reformulate the problem. Considering the example above, we divide by $\alpha$ and introduce
        an auxiliary variable $\lambda = (\beta / \alpha) h(v)$ where we take $\beta / \alpha \gg 1$, in order to demonstrate
        the ill-conditioned case. The system of equations then
        contains
        \begin{equation}
          f(v, \lambda) = g(v) + \lambda,
        \end{equation}
        and the auxiliary variable $\lambda$ is solved for in
        \begin{equation}
          h(v) - \frac{1}{\beta / \alpha} \lambda = 0,
        \end{equation}
        which is well-conditioned. An example approach of this type for nearly incompressible elasticity is known
        as the Herrmann formulation~\citep{Herrmann1965Elasticity}, where the pressure is introduced as an auxiliary variable to
        handle the large bulk modulus.
\end{enumerate}

On the other hand, \emph{inconsistent dimensions} indicate a mistake in the dimensional PDE. This is a
severe problem, often caused by incorrect dimensional transformation, wrong units of the
introduced quantities, or a more fundamental modeling mistake.

\paragraph{Example (Factorizing the Poisson equation).}

\begin{figure}[h]
  \centering
  \includegraphics{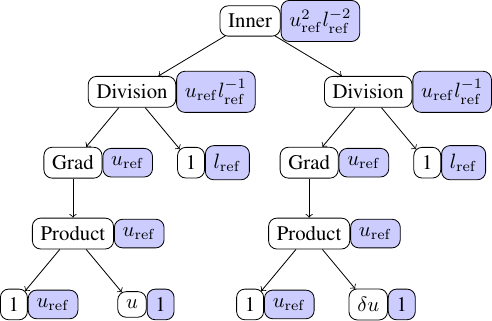}
  \caption{Quantity factorized DAG with dimensional scales annotated on each node. This
    representation exposes the dimensional homogeneity of the expression $\nabla u \cdot \nabla \delta u$
    with final units $u_\mathrm{ref}^2 l_\mathrm{ref}^{-2}$.}
  \label{fig:dag-factorization}
\end{figure}

This process is demonstrated in Fig.~\ref{fig:dag-factorization}. The diagram illustrates how the
factorizer traverses the DAG in a post-order manner (child nodes before parents), computing and
annotating each node with its corresponding dimensional expression. At terminal nodes (e.g.,
variables \ovalbox{$u$}, \ovalbox{$\delta u$}), the factorizer assigns appropriate dimensions based on their reference
quantities. As it traverses up the tree, it applies dimensional transformation rules for each operator:
multiplication combines dimensions via addition of dimension vectors, division subtracts dimension
vectors, and gradient operations incorporate the reference length scale $l_{\mathrm{ref}}$. The
final result at the root node represents the consolidated dimensional expression for the entire
bilinear form. In this case, $u_{\mathrm{ref}}^2 l_{\mathrm{ref}}^{-2}$ for the inner product of gradients.

More specifically, in the Poisson example the root factor is stored as the vector in $\mathbb Q^2$:
\begin{equation}
  \mathbf d(\ovalbox{Inner}) = (2, -2)^\trp
\end{equation}
assuming the provided ordering of dimensional quantities $(u_\mathrm{ref}, l_\mathrm{ref})$.

\paragraph{Example (Factorizing the Helmholtz equation).}

In order to demonstrate the handling of nonhomogeneous sums during factorization, we consider
the weak form of the Helmholtz equation:
\begin{equation}
  -\Delta u + \kappa^2 u = f,
\end{equation}
where $\kappa > 0$. The bilinear form corresponding to this equation is
\begin{equation}
  a(u, \delta u) = \int_\Omega \left( \nabla u \cdot \nabla \delta u + \kappa^2 u \, \delta u \right) \, \mathrm dx.
\end{equation}

\begin{figure}[h]
  \centering
  \includegraphics{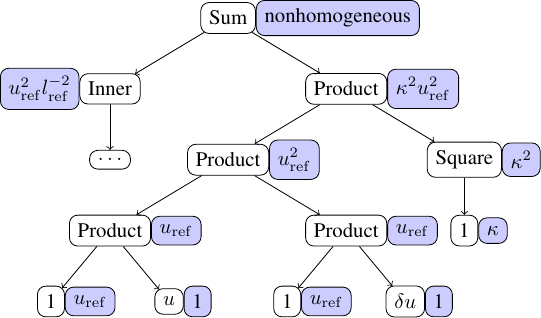}
  \caption{Factorized DAG for the nonhomogeneous sum $\nabla u \cdot \nabla \delta u + \kappa^2 u \delta u$. The sum node
    at the root shows inconsistency between its operands $u_\mathrm{ref}^2 l_\mathrm{ref}^{-2}$ and $\kappa^2 u_\mathrm{ref}^2$.}
  \label{fig:dag-factorization-helmholtz}
\end{figure}

The DAG of the integrand $\nabla u \cdot \nabla \delta u + \kappa^2 u \delta u$ after transformation and factorization steps
is shown in Fig.~\ref{fig:dag-factorization-helmholtz}, where we skip the detailed subgraph for the inner product of gradients
for brevity.

At this point, the factorizer detects that the \ovalbox{Sum} node is nonhomogeneous because its two
operands carry different scale factors:
\begin{equation}
  u_{\mathrm{ref}}^2 l_{\mathrm{ref}}^{-2} \neq \kappa^2 u_{\mathrm{ref}}^2.
\end{equation}
Firstly, it checks the dimensional consistency of the two factors. If they are dimensionally inconsistent,
i.e. $[u_{\mathrm{ref}}^2 l_{\mathrm{ref}}^{-2}] \neq [\kappa^2 u_{\mathrm{ref}}^2]$,
an exception is raised indicating a severe modeling error. It is clear from the above that
for the dimensional consistency to hold, $\kappa$ must have dimensions of inverse length,
$[\kappa] = \mathrm{L}^{-1}$.

More difficult is the case of consistent dimensions but inconsistent factors.
In this case, the factorizer raises an exception indicating that the sum is nonhomogeneous.
The user must then decide how to handle the inconsistent factors, e.g. by
letting $\kappa = l_\mathrm{ref}^{-1}$, as discussed earlier.

%%%%%%%%%%%%%%%%%%%%%%%%%%%%%%%%%%%%%%%%%%%%%%%%%%%%%%%%%%%%%%%%%%%%%%%%%%%%%%%

\subsection{Normalization}
\label{sec:normalization}

In this subsection, we describe the normalization step that produces dimensionless forms of
variational problems. This step is based on the factorization step described in the previous
subsection.

Let us assume that a function $F$ that describes the physical problem
can be written as a sum of $s$ terms $f_i$, each a function of the $n$ independent quantities
$\{q_1, \ldots, q_n\}$, i.e.,
\begin{equation}
  F(q_1,\dots,q_n) = f_1(q_1,\dots,q_n) + f_2(q_1,\dots,q_n) + \cdots + f_s(q_1,\dots,q_n).
  \label{eq:sum-function}
\end{equation}
This is clearly a very restrictive assumption, but many physical problems can be expressed in this
form, especially after linearization. The supporting assumption is that each term in the sum
comes from a different physical effect or mechanism, each with its own characteristic scale.
Typical examples include weak forms of PDEs, where each term in the bilinear or linear form
represents a different physical contribution (e.g., diffusion, convection, reaction, source terms,
etc.), energy functionals with multiple energy contributions, constrained optimization problems with
various penalty terms, multiphysics problems coupling different physical fields, etc.

Using the techniques explained in the previous subsection, the function is transformed and
factorized into
\begin{equation}
  F = \alpha_1 \widehat{f}_1 + \alpha_2 \widehat{f}_2 + \cdots + \alpha_s \widehat{f}_s,
\end{equation}
where $\widehat{f}_i$ are dimensionless (do not depend on any $q_i$) and $\alpha_i$ are
dimensional scaling factors that carry the units of each term $f_i$. The dimensional consistency
requires that all $\alpha_i$ have the same dimension, i.e., $[\alpha_i] = [\alpha_j]$ for all
$i, j = 1, \ldots, s$.
To nondimensionalize the expression, we divide through by a carefully chosen reference factor
$\alpha_\text{ref}$, which has the same dimension as the $\alpha_i$ factors. This yields a
dimensionless representation:
\begin{equation}
  \frac{F}{\alpha_\text{ref}} = \frac{\alpha_1}{\alpha_\text{ref}} \widehat{f}_1 + \frac{\alpha_2}{\alpha_\text{ref}} \widehat{f}_2
  + \cdots + \frac{\alpha_s}{\alpha_\text{ref}} \widehat{f}_s.
\end{equation}
The coefficients $\alpha_i / \alpha_\text{ref}$ are dimensionless numbers, ensuring that the entire
expression is invariant under transformations of the underlying unit system.
Using this approach, the dimensionless numbers (dimensionless groups) are defined uniquely,
and we obtain $s$ dimensionless groups for a sum of $s$ terms. Importantly, these dimensionless
groups are in general different from those obtained via Buckingham Pi analysis. If the groups
identified via normalization are not the same as those from Buckingham Pi analysis, this could indicate
that some of the terms $f_i$ are describing the same physical effect and could be combined, or
that important physical effects are not captured in the expression $F$, indicating their relevance
in boundary or initial conditions.

The reference factor $\alpha_\text{ref}$ is usually chosen as one of the $\alpha_i$ factors, based on modeling
requirements and the specific context of the problem. A common choice is to set
$\alpha_\text{ref} = \max_i |\alpha_i|$, i.e., as the most dominant term in the sum.

\paragraph{Example (Normalizing the Helmholtz equation).}

Continuing the Helmholtz equation example from the previous subsection, we first write the
bilinear form in the linear algebraic form:
\begin{equation}
  a(u, \delta u) = a_1(u, \delta u) + a_2(u, \delta u),
\end{equation}
where $a_1(u, \delta u) = \int_\Omega \nabla u \cdot \nabla \delta u \, \mathrm dx$ and
$a_2(u, \delta u) = \int_\Omega \kappa^2 u \, \delta u \, \mathrm dx$ represent the weak Laplacian
and mass contributions, respectively. After transformation and factorization (taking the topological dimension $d$), we obtain
\begin{equation}
  \begin{aligned}
    a & = u_{\mathrm{ref}}^2 l_{\mathrm{ref}}^{-2} l_{\mathrm{ref}}^d \widehat{a}_1
    + \kappa^2 u_{\mathrm{ref}}^2 l_{\mathrm{ref}}^d \widehat{a}_2.
  \end{aligned}
\end{equation}

The choice of reference factor is application-dependent, depending on the dominant physical effects, i.e.
if we expect diffusion or reaction to dominate the solution behavior. For an example here, we can choose
$\alpha_\text{ref} = u_{\mathrm{ref}}^2 l_{\mathrm{ref}}^{-2} l_{\mathrm{ref}}^d$, which yields the dimensionless form:
\begin{equation}
  \begin{aligned}
    \frac{a}{\alpha_\text{ref}} & = \widehat{a}_1 + \kappa^2 l_{\mathrm{ref}}^{2} \widehat{a}_2
  \end{aligned}
\end{equation}
which introduces the trivial dimensionless group $1$ and the dimensionless group $\kappa^2 l_{\mathrm{ref}}^{2}$. This dimensionless group
indicates the relative importance of the reaction term compared to the diffusion term in the Helmholtz
equation.

\subsection{Multiphysics problems}

The framework to handle multiphysics problems requires minor extensions to the normalization
step described in the previous subsection. In multiphysics problems, we typically have $m$ different
forms $F_1, F_2, \ldots, F_m$, each representing a physical equation with its own set of quantities
and own dimension. We proceed by transforming and factorizing each form separately, yielding
\begin{equation}
  \begin{aligned}
    F_1 & = \alpha_{1,1} \widehat{f}_{1,1} + \alpha_{1,2} \widehat{f}_{1,2} + \cdots + \alpha_{1,s_1} \widehat{f}_{1,s_1}, \\
    F_2 & = \alpha_{2,1} \widehat{f}_{2,1} + \alpha_{2,2} \widehat{f}_{2,2} + \cdots + \alpha_{2,s_2} \widehat{f}_{2,s_2}, \\
        & \vdots                                                                                                           \\
    F_m & = \alpha_{m,1} \widehat{f}_{m,1} + \alpha_{m,2} \widehat{f}_{m,2} + \cdots + \alpha_{m,s_m} \widehat{f}_{m,s_m}.
  \end{aligned}
\end{equation}

The choice of the reference factor is more involved in multiphysics problems, as we have the freedom
to choose a different reference factor for each form $F_i$. We denote these reference factors as
$\alpha_{\text{ref}, i}$ for $i = 1, \ldots, m$. Similar to the single-physics
case, the reference factors can be chosen based on modeling requirements, e.g., to highlight
dominant physical effects in each equation. An obvious choice is to set
$\alpha_{\text{ref}, i} = \max_j |\alpha_{i,j}|$, i.e., the most dominant term in each form $F_i$. This could,
however, neglect important coupling effects between different equations, so care must be taken in the choice of reference factors.

In the context of the weak form of linearized PDEs, i.e. when each $F_i$ is a bilinear or linear form, the individual
equations are typically summed together to form a coupled system. In order to satisfy the dimensional consistency across
this sum, it is required to choose the reference factors such that all normalized forms $F_i / \alpha_{\text{ref}, i}$
have the same dimension. This requirement ensures that the entire coupled system
\begin{equation}
  F = \frac{F_1}{\alpha_{\text{ref}, 1}} + \frac{F_2}{\alpha_{\text{ref}, 2}} + \cdots + \frac{F_m}{\alpha_{\text{ref}, m}}
\end{equation}
is dimensionally consistent.

%%%%%%%%%%%%%%%%%%%%%%%%%%%%%%%%%%%%%%%%%%%%%%%%%%%%%%%%%%%%%%%%%%%%%%%%%%%%%%%
%%%%%%%%%%%%%%%%%%%%%%%%%%%%%%%%%%%%%%%%%%%%%%%%%%%%%%%%%%%%%%%%%%%%%%%%%%%%%%%

\section{PDE nondimensionalization examples}

All examples in this section are available as demos in the GitHub repository
of the Dolfiny package, see \citet{Habera2025DolfinyRepo}.

\subsection{Incompressible Navier--Stokes}
The incompressible Navier--Stokes system is traditionally used as an example PDE to demonstrate
nondimensionalization and the appearance of dimensionless numbers. In its classical form we seek
velocity $\mathbf{v}$ and pressure $p$ such that
\begin{equation}
  \begin{aligned}
    \rho \frac{\partial \mathbf{v}}{\partial t}
    + \rho (\mathbf{v} \cdot \nabla) \mathbf{v}
    - \ddiv(2 \rho \nu \mathbf D(\mathbf{v}))
    + \nabla p
                     & = \rho \mathbf{g}, \\
    \ddiv \mathbf{v} & = 0.
  \end{aligned}
\end{equation}
Here, $\rho$ is the fluid density, $\nu$ the kinematic viscosity, $\mathbf D(\mathbf{v}) =
  \frac{1}{2} (\nabla\mathbf{v} + \nabla\mathbf{v}^\trp)$ the strain-rate tensor, and
$\mathbf{g}$ is an external body acceleration (e.g. gravity), so that $\rho \mathbf{g}$ is the body force per unit volume.

The Navier--Stokes system depends on seven dimensional quantities:
reference velocity $v_\mathrm{ref}$, reference pressure $p_\mathrm{ref}$, reference acceleration
$g_\mathrm{ref}$, reference time $t_\mathrm{ref}$, reference length $l_\mathrm{ref}$, density
$\rho$, and kinematic viscosity $\nu$. We can define these using the \texttt{Quantity} class,
see Fig.~\ref{fig:navier-stokes-quantities}. Example values are provided to demonstrate the dimensional analysis output.

\begin{figure}[h]
  \begin{minted}[fontsize=\footnotesize, linenos, frame=single]{python}
import sympy.physics.units as syu
from sympy import Symbol

from dolfiny.units import Quantity, buckingham_pi_analysis

nu    = Quantity(mesh, 1000, syu.millimeter**2 / syu.second, "nu")
rho   = Quantity(mesh, 5000, syu.kilogram / syu.m**3, "rho")
l_ref = Quantity(mesh, 1, syu.meter, "l_ref")
t_ref = Quantity(mesh, 1 / 60, syu.minute, "t_ref")
v_ref = Quantity(mesh, 1, syu.meter / syu.second, "v_ref")
p_ref = Quantity(mesh, 5000, syu.pascal, "p_ref")
g_ref = Quantity(mesh, 10, syu.meter / syu.second**2, "g_ref")

quantities = [v_ref, l_ref, rho, nu, g_ref, p_ref, t_ref]
buckingham_pi_analysis(quantities)
\end{minted}
  \caption{Definition of reference quantities for the Navier--Stokes system using the \texttt{Quantity} class. Each quantity includes its symbolic representation and corresponding SI
    units.}
  \label{fig:navier-stokes-quantities}
\end{figure}

The output of the above code snippet is shown in Fig.~\ref{fig:navier-stokes-pi-groups}.
Buckingham Pi analysis lists each quantity with its symbolic representation, value in base SI
units, the dimension matrix, and the resulting dimensionless groups. The dimensional matrix
is of size $7 \times 7$, since there are seven quantities and seven fundamental dimensions.
The ordering of columns and rows corresponds to the order of provided quantities and SI base dimensions,
respectively.

\begin{figure}[h]
  \centering
  \begin{subfigure}[b]{\textwidth}
    \centering
    \begin{tabular}{@{}lll@{}}
      \toprule
      \textbf{Symbol} & \textbf{Expression}                     & \textbf{Value (base units)}                  \\
      \midrule
      $v_\text{ref}$  & \SI{1}{\meter/\second}                  & \SI{1}{\meter/\second}                       \\
      $l_\text{ref}$  & \SI{1}{\meter}                          & \SI{1}{\meter}                               \\
      $\rho$          & \SI{5000}{\kilogram/\meter\cubed}       & \SI{5000}{\kilogram/\meter\cubed}            \\
      $\nu$           & \SI{1000}{\milli\meter\squared/\second} & \SI{0.001}{\meter\squared/\second}           \\
      $g_\text{ref}$  & \SI{10}{\meter/\second\squared}         & \SI{10}{\meter/\second\squared}              \\
      $p_\text{ref}$  & \SI{5000}{\pascal}                      & \SI{5000}{\kilogram/(\meter\second\squared)} \\
      $t_\text{ref}$  & \SI{0.01667}{\minute}                   & \SI{1}{\second}                              \\
      \bottomrule
    \end{tabular}
    \subcaption{Physical quantities}
  \end{subfigure}

  \vspace{1em}

  \begin{subfigure}[b]{\textwidth}
    \centering
    \begin{tabular}{@{}lccccccc@{}}
      \toprule
      \textbf{Dim.} & $v_\text{ref}$ & $l_\text{ref}$ & $\rho$ & $\nu$ & $g_\text{ref}$ & $p_\text{ref}$ & $t_\text{ref}$ \\
      \midrule
      L             & 1              & 1              & $-3$   & 2     & 1              & $-1$           & 0              \\
      M             & 0              & 0              & 1      & 0     & 0              & 1              & 0              \\
      T             & $-1$           & 0              & 0      & $-1$  & $-2$           & $-2$           & 1              \\
      \bottomrule
    \end{tabular}
    \subcaption{Dimension matrix (only non-zero rows shown: N, I, J, $\Theta$ all zero)}
  \end{subfigure}

  \vspace{1em}

  \begin{subfigure}[b]{\textwidth}
    \centering
    \begin{tabular}{@{}lll@{}}
      \toprule
      \textbf{Group} & \textbf{Expression}                        & \textbf{Value} \\
      \midrule
      $\Pi_1$        & $\nu/(l_\text{ref} v_\text{ref})$          & 0.001          \\
      $\Pi_2$        & $g_\text{ref} l_\text{ref}/v_\text{ref}^2$ & 10             \\
      $\Pi_3$        & $p_\text{ref}/(\rho v_\text{ref}^2)$       & 1              \\
      $\Pi_4$        & $t_\text{ref} v_\text{ref}/l_\text{ref}$   & 1              \\
      \bottomrule
    \end{tabular}
    \subcaption{Dimensionless groups}
  \end{subfigure}

  \caption{Buckingham Pi dimensional analysis for the Navier--Stokes system. This output was generated by the automated dimensional analysis framework presented in this work.}
  \label{fig:navier-stokes-pi-groups}
\end{figure}

The automated analysis identifies four dimensionless groups:
\begin{equation}
  \begin{aligned}
    \Pi_1 & = \frac{\nu}{l_\mathrm{ref} v_\mathrm{ref}} = \frac{1}{\mathrm{Re}}                & \quad \text{(inverse Reynolds number)},       \\
    \Pi_2 & = \frac{g_\mathrm{ref} l_\mathrm{ref}}{v_\mathrm{ref}^2} = \frac{1}{\mathrm{Fr}^2} & \quad \text{(squared inverse Froude number)}, \\
    \Pi_3 & = \frac{p_\mathrm{ref}}{\rho v_\mathrm{ref}^2} = \mathrm{Eu}                       & \quad \text{(Euler number)},                  \\
    \Pi_4 & = \frac{t_\mathrm{ref}}{l_\mathrm{ref} / v_\mathrm{ref}} = \frac{1}{\mathrm{St}}   & \quad \text{(inverse Strouhal number)}.
  \end{aligned}
\end{equation}

In fact, the dimensionless groups correspond to the classical dimensionless numbers used in fluid
mechanics. The Reynolds number characterizes the ratio of inertial to viscous forces, the Froude
number relates inertial to gravitational forces, the Euler number compares pressure forces to
inertial forces, and the Strouhal number describes oscillating flow dynamics. In this context,
however, the (inverse) Strouhal number characterizes the ratio of the reference time scale to the convective
time scale, so could also be called a global Courant-Friedrichs-Lewy (CFL) number.

The appearance of these classical dimensionless numbers through Buckingham Pi analysis
is more of a coincidence rather than a guarantee. As discussed previously, other independent
products of dimensionless groups can be formed by combining the identified groups.
The specific outcome of the groups in this example is due to the particular choice of
the algorithm used to compute a basis of the null space of the dimension matrix. We use
the SymPy implementation of the \texttt{Matrix.nullspace()} method, especially for matrices
over the rational field $\mathbb Q$, see \citet{sympy}. The algorithm tends to produce
basis vectors with small integer coefficients, which often correspond to classical
dimensionless numbers in physics and engineering. The only remaining variable is
the choice of ordering of the list of quantities provided.

So far, no PDE formulation has been considered in the dimensional analysis.
Let us now write the weak form of the incompressible Navier--Stokes equations.
The problem becomes: find $(\mathbf v, p) \in V \times P$ such that
\begin{equation}
  \int_\Omega \rho \frac{\partial\mathbf{v}}{\partial t} \cdot \delta \mathbf{v} \,\mathrm{dx}
  + \int_\Omega \rho (\mathbf{v} \cdot \nabla)\mathbf{v} \cdot \delta \mathbf{v} \,\mathrm{dx}
  + \int_\Omega 2 \rho \nu \mathbf D(\mathbf{v}) : \mathbf D(\delta \mathbf{v}) \, \mathrm{dx}
  - \int_\Omega p \ddiv \delta \mathbf{v} \,\mathrm{dx}
  - \int_\Omega \ddiv \mathbf{v} \delta p \,\mathrm{dx}
  = \int_\Omega \rho \mathbf{g} \cdot \delta \mathbf{v}\,\mathrm{dx}
  \label{eq:ns-weak-form}
\end{equation}
holds for all $(\delta \mathbf v, \delta p) \in V \times P$.
The matrix formulation of this nonlinear saddle-point system can be expressed as
\begin{equation}
  \begin{pmatrix}
    \mathcal{A}_t + \mathcal{A}_c + \mathcal{A}_v & \mathcal{B}^\trp \\
    \mathcal{B}                                   & \mathbf{0}
  \end{pmatrix}
  \begin{pmatrix}
    \mathbf{v} \\
    p
  \end{pmatrix}
  =
  \begin{pmatrix}
    \rho \mathbf{g} \\
    \mathbf{0}
  \end{pmatrix}
\end{equation}

where the operators are defined as
\begin{equation}
  \begin{aligned}
    (\mathcal{A}_t \mathbf{v}, \delta \mathbf{v}) & = \int_\Omega \rho\frac{\partial\mathbf{v}}{\partial t} \cdot \delta \mathbf{v}\,\mathrm{dx}, &
    (\mathcal{A}_c \mathbf{v}, \delta \mathbf{v}) & = \int_\Omega \rho(\mathbf{v} \cdot \nabla)\mathbf{v} \cdot \delta \mathbf{v}\,\mathrm{dx},     \\
    (\mathcal{A}_v \mathbf{v}, \delta \mathbf{v}) & = \int_\Omega 2 \rho \nu \mathbf D(\mathbf{v}) : \mathbf D(\delta \mathbf{v})\,\mathrm{dx},   &
    (\mathcal{B}\mathbf{v}, \delta p)             & = -\int_\Omega \ddiv \mathbf{v}\, \delta p \, \mathrm{dx},                                      \\
    (\mathcal{B}^\trp p, \delta \mathbf{v})       & = -\int_\Omega p \ddiv \delta \mathbf{v} \, \mathrm{dx}.
  \end{aligned}
\end{equation}

Going back to Eq. \eqref{eq:ns-weak-form} and Eq. \eqref{eq:sum-function}, the top-level
expression is a sum of six terms so we write
\begin{equation}
  \begin{aligned}
    F_\mathrm{NS}(v_\mathrm{ref}, p_\mathrm{ref}, g_\mathrm{ref}, t_\mathrm{ref}, l_\mathrm{ref}, \rho, \nu)
     & = \underbrace{(\mathcal{A}_t \mathbf{v}, \delta\mathbf{v})}_{\text{(unsteady)}} + \underbrace{(\mathcal{A}_c\mathbf{v}, \delta\mathbf{v})}_{\text{(convection)}} + \underbrace{(\mathcal{A}_v\mathbf{v}, \delta\mathbf{v})}_{\text{(viscous)}} \\
     & \quad + \underbrace{(\mathcal{B}\mathbf{v}, \delta p)}_{\text{(incompressibility)}} + \underbrace{(\mathcal{B}^\trp p, \delta\mathbf{v})}_{\text{(pressure)}} - \underbrace{(\rho \mathbf g, \delta\mathbf{v})}_{\text{(force)}}               \\
     & = 0.
  \end{aligned}
\end{equation}

For the dimensionless form of the equations, we have to proceed with transformation and factorization steps, see
subsections \ref{sec:transformation} and \ref{sec:factorization}. Since UFL does not have built-in support
for time derivatives, we discretize the time derivative using the backward Euler scheme:
\begin{equation}
  \frac{\partial \mathbf{v}}{\partial t} \approx \frac{\mathbf{v} - \mathbf{v}_0}{t_\mathrm{ref} \Delta t},
\end{equation}
where $\mathbf{v}_0$ is the velocity at the previous time step and $\Delta t$ the dimensionless time step size. The specific
choice of time discretization is not important for the dimensional analysis.

We introduce dimensional transformations as
\begin{equation}
  \begin{aligned}
    \mathbf{v} \mapsto v_\mathrm{ref}\, \mathbf{v}, \quad
    \mathbf{v}_0 \mapsto v_\mathrm{ref}\, \mathbf{v}_0, \quad
    p \mapsto p_\mathrm{ref}\, p, \quad
    \delta \mathbf{v} \mapsto v_\mathrm{ref}\, \delta\mathbf{v}, \quad
    \delta p \mapsto p_\mathrm{ref}\, \delta p
  \end{aligned}
\end{equation}
together with the transformation of spatial differential operators and integration measures
based on the characteristic length of the mesh, $l_\mathrm{ref}$. Introduction of the reference gravitational
acceleration using $g_\mathrm{ref}$ is done manually in the weak form. In the implementation, we define these transformations using a
dictionary that maps each UFL dimensionless node to its corresponding dimensional node as shown in
Fig.~\ref{fig:navier-stokes-mapping}.
\begin{figure}[h]
  \begin{minted}[fontsize=\footnotesize, linenos, frame=single]{python}
mapping = {
    mesh.ufl_domain(): l_ref,
    v:  v_ref * v,
    v0: v_ref * v0,
    p:  p_ref * p,
    δv: v_ref * δv,
    δp: p_ref * δp,
}
\end{minted}
  \caption{Dimensional transformation mapping for the Navier--Stokes weak form variables and
    operators. Each key-value pair associates a UFL object with its dimensional counterpart.}
  \label{fig:navier-stokes-mapping}
\end{figure}
It is important to notice that test functions $\delta \mathbf{v}$ and $\delta p$ are transformed too.
This is required for the dimensional consistency of the mixed weak form in Eq. \eqref{eq:ns-weak-form}.

In order to tag individual terms in the weak form, we use another dictionary to associate each
term with a descriptive name. This allows us to easily identify and manipulate the terms later in the
code.\footnote{The dictionary serves as an external tagging mechanism. UFL does
  not have a tagging mechanism where expressions or forms could be tagged and the tagged information
  would persist through consequent form manipulation.}

\begin{figure}[h]
  \begin{minted}[fontsize=\footnotesize, linenos, frame=single]{python}
from dolfiny.units import factorize, normalize

terms = {
    "unsteady":          ufl.inner(δv, rho * (v - v0) / (t_ref / n)) * ufl.dx,
    "convection":        ufl.inner(δv, rho * ufl.dot(v, ufl.grad(v))) * ufl.dx,
    "viscous":           ufl.inner(D(δv), 2 * rho * nu * D(v)) * ufl.dx,
    "incompressibility": ufl.inner(δp, ufl.div(v)) * ufl.dx,
    "pressure":         -ufl.inner(ufl.div(δv), p) * ufl.dx,
    "force":            -ufl.inner(δv, rho * g_ref * b) * ufl.dx,
}
factorized = factorize(terms, quantities, mode="factorize", mapping=mapping)

reference_term = "convection"
normalized = normalize(factorized, reference_term, quantities)
\end{minted}
  \caption{Code implementation for dimensional transformation and factorization of the Navier--Stokes
    weak form terms. The dictionary \texttt{terms} defines each component of the mixed weak
    formulation, while \texttt{mapping} establishes the dimensional transformations for all
    variables and operators.}
  \label{fig:navier-stokes-code}
\end{figure}

The dictionary definition for tagging terms, factorization, and normalization calls is shown
in Fig.~\ref{fig:navier-stokes-code}. The \texttt{factorize} function performs the dimensional transformation
and factorization of each term in the \texttt{terms} dictionary. The result is another dictionary called
\texttt{factorized} where each term is represented as a product of a dimensional factor and a
dimensionless UFL expression. The \texttt{normalize} function then divides each term by the
dimensional factor of the specified reference term. In this example we take the ``convection'' term.
This is the implementation of the process described in the subsection \ref{sec:normalization}.
The output of the dimensional factorization and normalization is shown in Fig.~\ref{fig:navier-stokes-terms}.
\begin{figure}[h]
  \centering
  \begin{subfigure}[b]{\textwidth}
    \centering
    \begin{tabular}{@{}ll@{}}
      \toprule
      \textbf{Expression}         & $l_\text{ref} \rho v_\text{ref}^3$       \\
      \textbf{Value (base units)} & \SI{5000}{\kilogram\meter/\second\cubed} \\
      \bottomrule
    \end{tabular}
    \subcaption{Reference factor from ``convection''}
  \end{subfigure}

  \vspace{0.5em}

  \begin{subfigure}[b]{\textwidth}
    \centering
    \begin{tabular}{@{}lll@{}}
      \toprule
      \textbf{Term}     & \textbf{Factor}                            & \textbf{Value} \\
      \midrule
      unsteady          & $l_\text{ref}/(t_\text{ref} v_\text{ref})$ & 1              \\
      convection        & $1$                                        & 1              \\
      viscous           & $\nu/(l_\text{ref} v_\text{ref})$          & 0.001          \\
      pressure          & $p_\text{ref}/(\rho v_\text{ref}^2)$       & 1              \\
      force             & $g_\text{ref} l_\text{ref}/v_\text{ref}^2$ & 10             \\
      incompressibility & $p_\text{ref}/(\rho v_\text{ref}^2)$       & 1              \\
      \bottomrule
    \end{tabular}
    \subcaption{Normalized terms}
  \end{subfigure}

  \caption{Automated dimensional factorization and normalization for the Navier--Stokes weak
    form. Each term is expressed with respect to the reference factor from the convection term,
    resulting in dimensionless coefficients. This output was generated by the dimensional analysis
    framework presented in this work.}
  \label{fig:navier-stokes-terms}
\end{figure}

Let us now walk through the steps to obtain the dimensionless form of the Navier--Stokes system in detail.
The transformed and factorized form reads (for two-dimensional case, $d=2$)
\begin{equation}
  \begin{aligned}
    F_\mathrm{NS} & = \rho v_\mathrm{ref}^2 t_\mathrm{ref}^{-1} l_\mathrm{ref}^{2} (\mathcal{A}_t\mathbf{v},\delta\mathbf{v})
    + \rho v_\mathrm{ref}^3 l_\mathrm{ref} (\mathcal{A}_c\mathbf{v},\delta\mathbf{v})
    + \rho \nu v_\mathrm{ref}^{2} (\mathcal{A}_v\mathbf{v},\delta\mathbf{v})                                                  \\
                  & \quad + p_\mathrm{ref} v_\mathrm{ref} l_\mathrm{ref} (\mathcal{B}^\trp p,\delta\mathbf{v})
    + p_\mathrm{ref} v_\mathrm{ref} l_\mathrm{ref} (\mathcal{B}\mathbf{v},\delta p)
    - \rho g_\mathrm{ref} v_\mathrm{ref} l_\mathrm{ref}^{2} (\mathbf{g},\delta\mathbf{v})                                     \\
                  & = 0,
  \end{aligned}
\end{equation}
and with the choice of normalization for convection dominated problems, i.e. setting
$\alpha_\mathrm{ref} = \rho v_\mathrm{ref}^3 l_\mathrm{ref}$ we obtain
\begin{equation}
  \begin{aligned}
    \frac{F_\mathrm{NS}}{\alpha_\mathrm{ref}} & = \frac{l_\mathrm{ref}}{t_\mathrm{ref} v_\mathrm{ref}} (\mathcal{A}_t\mathbf{v},\delta\mathbf{v})
    + (\mathcal{A}_c\mathbf{v},\delta\mathbf{v})
    + \frac{\nu}{v_\mathrm{ref} l_\mathrm{ref}} (\mathcal{A}_v\mathbf{v},\delta\mathbf{v})                                                        \\
                                              & \quad + \frac{p_\mathrm{ref}}{\rho v_\mathrm{ref}^2} (\mathcal{B}^\trp p,\delta\mathbf{v})
    + \frac{p_\mathrm{ref}}{\rho v_\mathrm{ref}^2} (\mathcal{B}\mathbf{v},\delta p)
    - \frac{g_\mathrm{ref} l_\mathrm{ref}}{v_\mathrm{ref}^2} (\mathbf{g},\delta\mathbf{v})                                                        \\
                                              & = 0.
  \end{aligned}
\end{equation}

Using the previously defined dimensionless numbers (Strouhal, Reynolds, Euler, and Froude) we can write the system
in the dimensionless matrix form:
\begin{equation}
  \begin{pmatrix}
    \mathrm{St} \mathcal{A}_t + \mathcal{A}_c + \frac{1}{\mathrm{Re}}\mathcal{A}_v & \mathrm{Eu} \mathcal{B}^\trp \\
    \mathrm{Eu} \mathcal{B}                                                        & \mathbf{0}
  \end{pmatrix}
  \begin{pmatrix}
    \mathbf v \\
    p
  \end{pmatrix}
  =
  \begin{pmatrix}
    \frac{1}{\mathrm{Fr}^2} \mathbf{g} \\
    \mathbf{0}
  \end{pmatrix}.
  \label{eq:ns-matrix-form-nondim}
\end{equation}

In addition to the nondimensionalization, a helper function \texttt{get\_dimension}
is provided. It can be used to fetch dimensions of arbitrary symbolic UFL expressions or forms.
This allows for sanity checks in the user code, as shown in Fig.~\ref{fig:dim-asserts-code}.

\begin{figure}[h]
  \begin{minted}[fontsize=\footnotesize, linenos, frame=single]{python}
from dolfiny.units import get_dimension

dimsys = syu.si.SI.get_dimension_system()
assert dimsys.equivalent_dims(get_dimension(D(v), quantities, mapping), 1 / syu.time)
  \end{minted}
  \caption{Example of using \texttt{dolfiny.units.get\_dimension} to check the dimensions of UFL
    expressions. The assertion checks that the dimension of the strain--rate tensor
    $[\mathbf D(u)] = \mathrm{T}^{-1}$. Equivalency of dimensions is verified using SymPy's SI
    dimension system.}
  \label{fig:dim-asserts-code}
\end{figure}

\paragraph{Conditioning of the Navier--Stokes system.}

Let us now explore the choice of the reference quantities in the Navier--Stokes system. The main
motivation is to find a set of reference quantities that lead to a well-conditioned linear system in
the dimensionless form, Eq. \eqref{eq:ns-matrix-form-nondim}. There are seven dimensional quantities in
the $F_\mathrm{NS}$ system, but two of them, density $\rho$ and kinematic viscosity $\nu$, are material
properties and cannot be chosen freely. In the choice of $v_\mathrm{ref}$ and $p_\mathrm{ref}$ we
have the complete freedom, i.e. any positive values will be factored out of the system and the
dimensional velocity $v_\mathrm{ref} \mathbf v$ and dimensional pressure $p_\mathrm{ref} p$
will be independent of the choice of $v_\mathrm{ref}$ and $p_\mathrm{ref}$.

For the sake of simplicity, we'll assume that the linearized and discretized Navier--Stokes system
leads to a matrix problem of the form
\begin{equation}
  \underbrace{
    \begin{pmatrix}
      \mathbf{A}               & \mathrm{Eu} \, \mathbf B^\trp \\
      \mathrm{Eu} \, \mathbf B & 0
    \end{pmatrix}}_{\mathbf K(\mathrm{Eu})}
  \begin{pmatrix}
    \mathbf{v} \\
    p
  \end{pmatrix}
  =
  \begin{pmatrix}
    \frac{1}{\mathrm{Fr}^2} \mathbf{g} \\
    \mathbf{0}
  \end{pmatrix}
  \label{eq:ns-saddle-point-system}
\end{equation}

where $\mathbf{A}$ is the linearized and discretized operator for the unsteady, convection and viscous terms. For
the purpose of this example we'll examine the conditioning of the matrix
$\mathbf K(\mathrm{Eu}) \in \mathbb R^{N \times N}$ as a function of the dimensionless Euler number $\mathrm{Eu}$.
The matrix has $N = n + m$ rows and columns, where $n$ is the number of velocity degrees of freedom
and $m$ the number of pressure degrees of freedom.
We'll assume that the matrix
$\mathbf{A}$ is provided and omit the discussion of the dependence of its spectrum on the Strouhal and
Reynolds numbers. This is an oversimplification, but it allows us to focus on the influence
of the Euler number on the conditioning, as a prototype for other saddle-point systems.

\begin{figure}[h]
  \centering
  \includegraphics{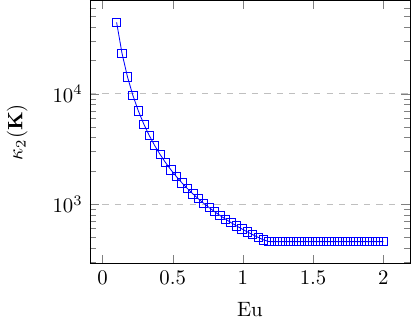}
  \caption{Condition number $\kappa_2$ of the Navier--Stokes matrix $\mathbf{K}(\mathrm{Eu})$ as a
    function of the Euler number. The plot demonstrates how the choice of pressure reference scale
    $p_\mathrm{ref}$ affects the numerical conditioning of the linearized system. The condition number
    was measured using NumPy's \texttt{linalg.cond} resulting from a $10 \times 10$ triangular mesh and Taylor-Hood
    $\mathbf P_2 - P_1$ discretization.}
  \label{fig:euler-conditioning}
\end{figure}

As can be seen in Fig.~\ref{fig:euler-conditioning}, the 2-norm condition number $\kappa_2(\mathbf K)$ of
the Navier--Stokes matrix $\mathbf K(\mathrm{Eu})$ is highly sensitive to the choice of the pressure
reference scale $p_\mathrm{ref}$, which is controllable through the dimensionless Euler number
$\mathrm{Eu} = p_\mathrm{ref} / (\rho v_\mathrm{ref}^2)$.
The naive choice of $p_\mathrm{ref} = \SI{1}{\pascal}$ (and in turn $\mathrm{Eu} = 0.0002$)
leads to an ill-conditioned system. In Appendix \ref{ap:ns-condition-number}, we show that
there exists a constant $C > 0$ independent of $\mathrm{Eu}$ such that
\begin{equation}
  \kappa_2(\mathbf K) \geq C \, \mathrm{Eu}^{-2m/N}.
\end{equation}

In other words, as the Euler number $\mathrm{Eu} \to 0$, the condition number $\kappa_2(\mathbf K) \to \infty$.
This is clearly visible in Fig.~\ref{fig:euler-conditioning}. The result is not surprising, since
the Euler number scales the off-diagonal block $\mathbf B$, and this block is essential for the
invertibility of the saddle-point system. In addition, for values of $\mathrm{Eu} > 1.2$ the condition
number stabilizes and remains nearly constant. This happens because the smallest singular value
$\sigma_1$ of the matrix $\mathbf K$ becomes determined by the velocity block $\mathbf A$ rather
than the off-diagonal incompressibility block $\mathbf B$.

With these observations, we can conclude that the reference pressure scale in this example should be
made sufficiently large, i.e. $p_\mathrm{ref} > \SI{5000}{\pascal}$ (such that $\mathrm{Eu} > 1$),
to ensure numerical stability and conditioning of the linearized Navier--Stokes system.

\paragraph{Relation to matrix preconditioning.}

In more general terms, the nondimensionalization can be seen as a simple diagonal preconditioning
technique that improves the conditioning of the resulting linear systems. This is particularly
important when solving multiphysics problems where different physical processes interact.

Consider the algebraic system of the linearized Navier--Stokes equations corresponding to the
dimensional weak form (before applying transformations of $\mathbf v$ and $p$ and before normalization):
\begin{equation}
  \underbrace{\begin{pmatrix}
      \rho t_\mathrm{ref}^{-1} l_\mathrm{ref}^{2} \mathbf{A}_t
      + \rho v_\mathrm{ref} l_\mathrm{ref} \mathbf{A}_c
      + \rho \nu \mathbf{A}_v  & l_\mathrm{ref} \mathbf B^\trp \\
      l_\mathrm{ref} \mathbf B & \mathbf{0}
    \end{pmatrix}
  }_{\widetilde{\mathbf K}}
  \begin{pmatrix}
    \mathbf v \\
    p
  \end{pmatrix}
  =
  \begin{pmatrix}
    \rho g_\mathrm{ref} l_\mathrm{ref}^{2} \mathbf g \\
    \mathbf{0}
  \end{pmatrix}.
\end{equation}

The nondimensionalization process is mathematically equivalent to applying a two-sided (left and
right) diagonal preconditioner $\mathbf{K}= \mathbf{D}_l \widetilde{\mathbf{K}} \mathbf{D}_r$. The
transformations are:
\begin{enumerate}
  \item \emph{Right preconditioning} (transformation of variables). Substituting dimensional
        variables with scaled dimensionless ones ($\mathbf{v} \mapsto v_\mathrm{ref} \mathbf{v}$, $p \mapsto
          p_\mathrm{ref} p$) is equivalent to right-multiplying the system by
        \begin{equation}
          \mathbf D_r =
          \begin{pmatrix}
            v_\mathrm{ref} \mathbf I & 0                        \\
            0                        & p_\mathrm{ref} \mathbf I
          \end{pmatrix}.
        \end{equation}
  \item \emph{Left preconditioning} (test function scaling and normalization). Scaling the test
        functions ($\delta \mathbf{v} \mapsto v_\mathrm{ref} \delta \mathbf{v}, \delta p \mapsto p_\mathrm{ref} \delta p$) and dividing the entire
        system by the reference scale ($\alpha_\mathrm{ref} = \rho v_\mathrm{ref}^3
          l_\mathrm{ref}$) is equivalent to left-multiplying by
        \begin{equation}
          \mathbf D_l = \frac{1}{\rho v_\mathrm{ref}^3 l_\mathrm{ref}}
          \begin{pmatrix}
            v_\mathrm{ref} \mathbf I & 0                        \\
            0                        & p_\mathrm{ref} \mathbf I
          \end{pmatrix}.
        \end{equation}
\end{enumerate}

The left and right preconditioned system matrix is
equivalent to the dimensionless system in Eq. \eqref{eq:ns-matrix-form-nondim}. From a numerical
linear algebra perspective, this procedure is a form of \emph{matrix equilibration}
\citep{vanderSluis1969Equilibration}. However, unlike algebraic equilibration which scales blindly based
on matrix entry magnitudes, nondimensionalization scales based on physical characteristic units.
This physics-aware scaling is robust against numerical noise and helps cluster eigenvalues, thereby
improving the convergence rate of Krylov subspace solvers (e.g., GMRES) for multiphysics problems.
The importance of dimensional consistency in preconditioning saddle-point systems has been further
explored by \citet{herzog2021dimensionally}, who demonstrated that aligning preconditioner parameters
with natural physical units yields robust performance across varying problem parameters.

\paragraph{Relation to Full Operator Preconditioning.}

It is crucial to distinguish between algebraic preconditioning and the operator-level transformation
presented here. Recent work by \citet{Mohr2024FOP} introduces the concept of \emph{Full
  Operator Preconditioning} (FOP) to describe techniques that modify the underlying partial
differential operator prior to discretization, contrasting them with standard preconditioning
which operates on the already assembled matrix.

While the nondimensionalization described in the previous section is algebraically equivalent to
diagonal scaling $\mathbf{K} = \mathbf{D}_l \widetilde{\mathbf{K}} \mathbf{D}_r$, the numerical
realization differs fundamentally. Standard preconditioning assembles the ill-conditioned matrix
$\widetilde{\mathbf{K}}$ first, and subsequently applies $\mathbf{D}_l$ and $\mathbf{D}_r$ during the iterative
solve. In contrast, our framework applies the scaling weights directly to the variational form
\emph{before} assembly. We assemble $\mathbf{K}$ directly, effectively commuting the
operations of discretization and scaling.

This distinction has significant implications for floating-point accuracy. As demonstrated by
\citet{Mohr2024FOP}, assembling $\widetilde{\mathbf{K}}$ involving quantities with disparate magnitudes
(e.g., $1$ and $10^{-15}$) can lead to catastrophic cancellation or loss of significant digits in
finite precision arithmetic. Algebraic preconditioning cannot recover this lost information. By
enforcing unit consistency at the UFL level, our framework implements the simplest form of the
Full Operator Preconditioning,
ensuring that the assembly process operates on $\mathcal{O}(1)$ quantities, thereby preserving the
numerical accuracy of the discrete operator. The effect of the \emph{row-column equilibration}
on the accuracy of the solution is also noted in \citet[\S 3.5.2]{Golub2013Matrix}.
The accuracy as the driving principle for nondimensionalization is further discussed
in the following section on hyperelasticity.

%%%%%%%%%%%%%%%%%%%%%%%%%%%%%%%%%%%%%%%%%%%%%%%%%%%%%%%%%%%%%%%%%%%%%%%%%%%%%%%

\subsection{Neo-Hooke hyperelasticity}

A displacement $\mathbf u \in U_0$ of a hyperelastic material solves
\begin{equation}
  \min_{u \in U_0} \int_\Omega \tilde W(\mathbf u) \, \mathrm dx - W_\text{ext}(\mathbf u),
\end{equation}
where $\tilde{W}: U_0 \to \mathbb R$ is the strain energy density function. One common form of the
compressible Neo-Hooke material model
\begin{equation}
  \tilde W = \tilde W_\text{shear} + \tilde W_\text{bulk} =
  \underbrace{\frac{\mu}{2} \left( I_1 - 3 - 2 \ln J \right)}_{\text{shear term}}
  + \underbrace{\frac{\kappa}{2} \left( J - 1 \right)^2}_{\text{bulk term}},
\end{equation}
is defined in \citet[Eq. (1.4)]{Pence2014NeoHookean}. Here, $I_1 = \mathrm{tr}(\mathbf C)$ is the first invariant of the right Cauchy--Green deformation
tensor $\mathbf C = \mathbf F^\trp \mathbf F$ with the deformation gradient $\mathbf F = \mathbf I + \nabla \mathbf u$. Here,
$J = \sqrt{\det \mathbf C} = \abs{\det \mathbf F}$ is the Jacobian of the deformation,
and $\mu$ and $\kappa$ the shear and bulk moduli, respectively. External work done by
traction on the boundary $\Gamma_N$ is given by
\begin{equation}
  W_\text{ext} = \int_{\Gamma_N} \mathbf t \cdot \mathbf u \, \mathrm{d}s
\end{equation}
where $\mathbf t$ is the boundary traction.

In this example, we restrict ourselves to three-dimensional problems, $d=3$.
The Neo-Hooke hyperelasticity problem depends on five dimensional quantities:
reference displacement scale $u_\mathrm{ref}$, reference traction $\tau_\mathrm{ref}$,
reference length $l_\mathrm{ref}$, shear modulus $\mu$, and bulk modulus $\kappa$.
These can be introduced using the \texttt{Quantity} class as shown in Fig.~\ref{fig:neo-hooke-quantities}.
In the code, Poisson's ratio $\nu$, Young's modulus $E$, and Lamé's first parameter $\lambda$ are auxiliary
quantities used to define $\mu$ and $\kappa$ so are not included in the dimensional analysis.

\begin{figure}[h]
  \begin{minted}[fontsize=\footnotesize, linenos, frame=single]{python}
import sympy.physics.units as syu
from sympy import Symbol

from dolfiny.units import Quantity, buckingham_pi_analysis

nu = 0.4
E = Quantity(mesh, 2, syu.giga * syu.pascal, "E")
λ = Quantity(mesh, E.scale * nu / ((1 + nu) * (1 - 2 * nu)), E.unit, "λ")

μ = Quantity(mesh, E.scale / (2 * (1 + nu)), E.unit, "μ")
κ = Quantity(mesh, λ.scale + 2 / 3 * μ.scale, E.unit, "κ")

l_ref = Quantity(mesh, 1.0, syu.millimeter, "l_ref")
τ_ref = Quantity(mesh, 100, syu.kilo * syu.pascal, "τ_ref")
u_ref = Quantity(mesh, 0.001, syu.millimeter, "u_ref")

quantities = [μ, κ, l_ref, τ_ref, u_ref]
buckingham_pi_analysis(quantities)
\end{minted}
  \caption{Definition of reference quantities for the Neo-Hooke hyperelasticity system using the
    \texttt{Quantity} class. Each quantity includes its symbolic representation and corresponding SI
    units.}
  \label{fig:neo-hooke-quantities}
\end{figure}

\begin{figure}[h]
  \centering
  \begin{subfigure}[b]{\textwidth}
    \centering
    \begin{tabular}{@{}lll@{}}
      \toprule
      \textbf{Symbol}   & \textbf{Expression}      & \textbf{Value (base units)}                     \\
      \midrule
      $\mu$             & \SI{7.143e8}{\pascal}    & \SI{7.143e8}{\kilogram/(\meter\second\squared)} \\
      $\kappa$          & \SI{3.333e9}{\pascal}    & \SI{3.333e9}{\kilogram/(\meter\second\squared)} \\
      $l_\text{ref}$    & \SI{1}{\milli\meter}     & \SI{0.001}{\meter}                              \\
      $\tau_\text{ref}$ & \SI{1e5}{\pascal}        & \SI{1e5}{\kilogram/(\meter\second\squared)}     \\
      $u_\text{ref}$    & \SI{0.001}{\milli\meter} & \SI{1e-6}{\meter}                               \\
      \bottomrule
    \end{tabular}
    \subcaption{Physical quantities}
  \end{subfigure}

  \vspace{1em}

  \begin{subfigure}[b]{\textwidth}
    \centering
    \begin{tabular}{@{}lccccc@{}}
      \toprule
      \textbf{Dim.} & $\mu$ & $\kappa$ & $l_\text{ref}$ & $\tau_\text{ref}$ & $u_\text{ref}$ \\
      \midrule
      L             & $-1$  & $-1$     & 1              & $-1$              & 1              \\
      M             & 1     & 1        & 0              & 1                 & 0              \\
      T             & $-2$  & $-2$     & 0              & $-2$              & 0              \\
      \bottomrule
    \end{tabular}
    \subcaption{Dimension matrix (only non-zero rows shown: N, I, J, $\Theta$ all zero)}
  \end{subfigure}

  \vspace{1em}

  \begin{subfigure}[b]{\textwidth}
    \centering
    \begin{tabular}{@{}lll@{}}
      \toprule
      \textbf{Group} & \textbf{Expression}         & \textbf{Value} \\
      \midrule
      $\Pi_1$        & $\kappa/\mu$                & 4.67           \\
      $\Pi_2$        & $\tau_\text{ref}/\mu$       & 0.00014        \\
      $\Pi_3$        & $u_\text{ref}/l_\text{ref}$ & 0.001          \\
      \bottomrule
    \end{tabular}
    \subcaption{Dimensionless groups}
  \end{subfigure}

  \caption{Buckingham Pi dimensional analysis for the Neo-Hooke hyperelasticity system. This output was generated by the automated dimensional analysis framework presented in this work.}
  \label{fig:neo-hooke-pi-groups}
\end{figure}

The output of the Buckingham Pi dimensional analysis is shown in Fig.~\ref{fig:neo-hooke-pi-groups}.
The automated analysis identifies three dimensionless groups:
\begin{equation}
  \begin{aligned}
    \Pi_1 & = \frac{\kappa}{\mu}                    & \quad \text{(bulk to shear ratio)},          \\
    \Pi_2 & = \frac{\tau_\mathrm{ref}}{\mu}         & \quad \text{(traction to shear ratio)},      \\
    \Pi_3 & = \frac{u_\mathrm{ref}}{l_\mathrm{ref}} & \quad \text{(displacement to length ratio)}.
  \end{aligned}
\end{equation}

The first Pi group $\Pi_1$ expresses the ratio of bulk to shear modulus, which is related to
the relative contribution of the respective energy terms in the strain energy density function. The second
group $\Pi_2$ relates the reference traction to the shear modulus, indicating the relative magnitude
of external loading compared to the material's shear resistance. The third group $\Pi_3$ compares
the reference displacement to the reference length, providing a measure of the deformation scale
relative to the characteristic size of the domain.

For a hyperelastic material, the second Piola-Kirchhoff stress tensor $\mathbf S$ is given by
\begin{equation}
  \mathbf S =  \frac{\partial \tilde W}{\partial \mathbf E} = 2 \frac{\partial \tilde W}{\partial \mathbf C}.
\end{equation}
The weak form reads: find $\mathbf u \in U_0$ such that
\begin{equation}
  - \frac{1}{2} \int_\Omega \mathbf S : \frac{\partial \mathbf C}{\partial \mathbf u}[\delta \mathbf u] \, \mathrm{d}x
  + \int_{\Gamma_N} \mathbf t \cdot \delta \mathbf u \, \mathrm{d}s = 0
\end{equation}
holds for all $\delta \mathbf u \in U_0$, where $\frac{\partial \mathbf C}{\partial \mathbf u}[\delta \mathbf u]$ is the G\^ateaux derivative
of the right Cauchy-Green deformation tensor in the direction of $\delta \mathbf u$.

Mapping that introduces dimensional transformations is chosen as
\begin{equation}
  \begin{aligned}
    \mathbf u \mapsto u_\mathrm{ref} \, \mathbf u,
    \quad \delta \mathbf u \mapsto u_\mathrm{ref} \, \delta \mathbf u,
  \end{aligned}
\end{equation}
and is complemented by the spatial scaling of the differential operators and integration measures
with the reference length $l_\mathrm{ref}$. Introduction of the scaling for the traction force based
on the reference traction $\tau_\mathrm{ref}$ is done manually in the weak form.

The dimensional factorization attempt correctly identifies a potential issue and raises an exception. The exception message reads:
\begin{minted}[fontsize=\footnotesize, frame=single]{text}
  RuntimeError: Inconsistent factors
     I
---> +
     { A | A_{i_{25}, i_{26}} = (grad({ A | A_{i_{22}} = ... }.
Different factors: 1 != u_ref/l_ref.
\end{minted}

The factorization process detects an inconsistency of factors, as described in subsection \ref{sec:factorization}. Specifically,
the deformation gradient cannot be factorized homogeneously in terms of the
reference quantities $u_\mathrm{ref}$ and $l_\mathrm{ref}$. The gradient operator introduces a division
by the length scale $l_\mathrm{ref}$, while the displacement field $\mathbf u$ is scaled by the
displacement scale $u_\mathrm{ref}$. This produces the transformed expression:
\begin{equation}
  \mathbf F = \mathbf I + \frac{u_\mathrm{ref}}{l_\mathrm{ref}} \nabla \mathbf u = \mathbf I + \Pi_3 \nabla \mathbf u.
\end{equation}
We add two quantities of \emph{potentially} vastly different magnitudes. One possible solution
is to choose the reference displacement scale to match the reference length scale, $u_\mathrm{ref} = l_\mathrm{ref}$,
so that both terms in the sum are of the same order of magnitude. However, this choice is only valid if
the expected displacements are of the same order as the domain size, which is not the case for small
deformations.

The inconsistency of factors could manifest as numerical instability in the solution process,
especially when the dimensionless group $\Pi_3 = u_\mathrm{ref} / l_\mathrm{ref}$ is very small or very large.
To demonstrate this, consider the setup of $\Pi_3 = 10^{-k}$, which for large $k$ is a limit of small
deformations. Evaluation of the deformation gradient leads to losing $k$ digits of accuracy.
In an extreme case of $\Pi_3 = 10^{-16}$, the deformation gradient evaluates to identity matrix
in IEEE 754 double precision arithmetic, losing all information about the deformation. In contrast,
starting from the linearized strain energy density function and introducing the Green-Lagrange strain
\begin{equation}
  \mathbf E = \frac{1}{2} (\mathbf C - \mathbf I) = \frac{1}{2} (\mathbf F^\trp \mathbf F - \mathbf I),
\end{equation}
allows us to rescale the displacement field without losing the relative accuracy. The issue of numerical
stability in hyperelastic material models is discussed in more detail in \citet{Shakeri2024Stablenumerics}.

\paragraph{Reformulation using Green-Lagrange strain split.}

The strain energy density function can be reformulated in terms of the Green-Lagrange strain $\mathbf E$,
see Appendix \ref{ap:neo-hooke-expansion}, as
\begin{equation}
  \begin{aligned}
    \tilde W & = \mu \left( \tr(\mathbf E) - \ln J \right) + \frac{\kappa}{2} (J - 1)^2, \\
    J        & = \sqrt{\det(\mathbf I + 2\mathbf E)}.
  \end{aligned}
\end{equation}

This form is not suitable for dimensional factorization either, since the determinant term $J$ still
contains expressions of form $1 + \mathcal O(\Pi_3)$. However, we can expand the strain energy density function
in terms of the dimensionless group $\Pi_3$ assuming small deformations, i.e.,
\begin{equation}
  \tilde W = \tilde W^{(2)} + \tilde W^{(3)} + \mathcal{O}(\Pi_3^4),
\end{equation}
where $\tilde W^{(2)}$ and $\tilde W^{(3)}$ are the second- and third-order terms in the expansion, respectively.
In order to achieve the homogeneity in the dimensional quantities $\mu$, $\kappa$, $l_\mathrm{ref}$, and $u_\mathrm{ref}$,
we express the linearized Neo-Hooke strain energy density as
\begin{equation}
  \begin{aligned}
    \tilde W & = \underbrace{\mu \tr(\mathbf E_1^2)}_{\tilde W^{(2)}_\mu} + \underbrace{\frac{\kappa}{2} \tr(\mathbf E_1)^2}_{\tilde W^{(2)}_\kappa} \\
             & + \underbrace{\mu \Bigl( 2 \mathbf E_1 : \mathbf E_2 - \frac{4}{3} \tr(\mathbf E_1^3)\Bigr)}_{\tilde W^{(3)}_\mu} +
    \underbrace{\kappa \Bigl( \tr(\mathbf E_1) \tr(\mathbf E_2) + \frac{1}{2} \tr(\mathbf E_1)^3 - \tr(\mathbf E_1) \tr(\mathbf E_1^2) \Bigr)}_{\tilde W^{(3)}_\kappa}
    + \mathcal{O}(\norm{\mathbf E}^4),
  \end{aligned}
\end{equation}
where $\mathbf E_1 = \frac{1}{2} (\nabla \mathbf u + \nabla \mathbf u^\trp)$ is the linearized strain tensor
and $\mathbf E_2 = \frac{1}{2} \nabla \mathbf u^\trp \nabla \mathbf u$ is the second-order nonlinear correction.
The reason for this split is that all four terms are now homogeneous in the dimensional quantities, and
they are grouped to represent the second- and third-order contributions in terms of the dimensionless group $\Pi_3$ and
the bulk and shear effects.

The definition of the weak form terms and their dimensional transformation follows
the same pattern as in the previous examples, see Fig.~\ref{fig:navier-stokes-code} for reference.
We define four terms ``shear\_2'', ``bulk\_2'', ``shear\_3'', and ``bulk\_3''
corresponding to the second- and third-order contributions from shear and bulk parts of the strain
energy density function, respectively. The external force term is defined as ``force''.

\begin{figure}[h]
  \centering
  \begin{subfigure}[b]{\textwidth}
    \centering
    \begin{tabular}{@{}ll@{}}
      \toprule
      \textbf{Expression}         & $l_\text{ref} u_\text{ref}^2 \kappa$                   \\
      \textbf{Value (base units)} & \SI{3.333e-6}{\kilogram\meter\squared/\second\squared} \\
      \bottomrule
    \end{tabular}
    \subcaption{Reference factor from ``bulk\_2''}
  \end{subfigure}

  \vspace{0.5em}

  \begin{subfigure}[b]{\textwidth}
    \centering
    \begin{tabular}{@{}lll@{}}
      \toprule
      \textbf{Term} & \textbf{Factor}                                      & \textbf{Value} \\
      \midrule
      shear\_2      & $\mu/\kappa$                                         & 0.2143         \\
      bulk\_2       & $1$                                                  & 1              \\
      shear\_3      & $u_\text{ref} \mu/(l_\text{ref} \kappa)$             & 0.0002143      \\
      bulk\_3       & $u_\text{ref}/l_\text{ref}$                          & 0.001          \\
      force         & $l_\text{ref} \tau_\text{ref}/(u_\text{ref} \kappa)$ & 0.03           \\
      \bottomrule
    \end{tabular}
    \subcaption{Normalized terms}
  \end{subfigure}

  \caption{Automated dimensional factorization and normalization for the Neo-Hooke
    hyperelasticity weak form. Each term is expressed with respect to the reference factor from
    the bulk term, resulting in dimensionless coefficients. This output was generated by the dimensional analysis
    framework presented in this work.}
  \label{fig:neo-hooke-terms}
\end{figure}

The result of the dimensional factorization and normalization is shown in Fig.~\ref{fig:neo-hooke-terms}.
The reference factor is chosen from the bulk second-order term. After
normalization, we observe the appearance of dimensionless coefficients corresponding to the
previously identified $\Pi$ groups. For the example values of the reference quantities
defined in Fig.~\ref{fig:neo-hooke-quantities},
we observe that the shear contribution is approximately $21\%$ of
the bulk contribution, indicating a material behavior dominated by volumetric effects. The
third-order contributions are significantly smaller, reflecting the small deformation assumption
inherent in the expansion. The external force term is also relatively small, suggesting that the
applied traction is moderate compared to the material's bulk stiffness.

Dimensionless form of the Neo-Hooke hyperelasticity is summarized in Appendix \ref{ap:neo-hooke-dimless}.

%%%%%%%%%%%%%%%%%%%%%%%%%%%%%%%%%%%%%%%%%%%%%%%%%%%%%%%%%%%%%%%%%%%%%%%%%%%%%%%

\subsection{Poisson--Nernst--Planck}

The following demonstrates the use of the dimensional analysis framework for the
multiphysics Poisson--Nernst--Planck (PNP) equations. The PNP system models the transport of charged
species under the influence of an electric field. It consists of the Nernst--Planck model for the flux of
ion concentrations coupled with the Poisson equation for the electric potential. The PNP model is taken from
\citet{Samson1999PNP}.

There are $N$ equations of mass conservation for the $N$ ion concentrations $c_i$:
\begin{equation}
  \frac{\partial c_i}{\partial t} + \ddiv \mathbf j_i = 0, \quad i = 1, \ldots, N,
  \label{eq:pnp-nernst-planck}
\end{equation}
where the flux $\mathbf j_i$ is given by the Nernst--Planck equation:
\begin{equation}
  \mathbf j_i = -D_i \left( \nabla c_i + \frac{z_i F}{R T} c_i \nabla \varphi + c_i \nabla \ln(\gamma_i) \right).
\end{equation}
Here $c_i$ is the concentration of species $i$ (in $\mathrm{mol}/\mathrm{m}^3$), $D_i$ is the diffusion coefficient
(in $\mathrm{m}^2/\mathrm{s}$), $z_i$ is the valence (dimensionless), $F = \SI{9.6485e4}{\coulomb/\mol}$ is the Faraday constant,
$R = \SI{8.3145}{\joule/(\mol\kelvin)}$ is the universal gas constant, $T$ is the absolute temperature (in $\mathrm{K}$),
$\varphi$ is the electric potential (in $\mathrm{V}$), and $\gamma_i$ is the activity coefficient (dimensionless).

The extended Debye--Hückel model is given by
\begin{equation}
  \ln(\gamma_i) = - \frac{A z_i^2 \sqrt{I}}{1 + B a_i \sqrt{I}},
  \label{eq:pnp-debye-huckel}
\end{equation}
where $A$ and $B$ are the temperature-dependent Debye--Hückel constants computed from
\begin{equation}
  A = \frac{\sqrt{2} F^2 e_0}{8 \pi (\epsilon R T)^{3/2}}, \quad B = \sqrt{\frac{2 F^2}{\epsilon R T}},
\end{equation}
and $e_0 = \SI{1.6022e-19}{\coulomb}$ is the elementary charge, $\epsilon = \epsilon_0 \epsilon_r$ is the permittivity
of the medium computed from the dimensionless relative permittivity $\epsilon_r$ and the permittivity of vacuum $\epsilon_0$
(in $\mathrm{F}/\mathrm{m}$) and $a_i$ is the radius of the ion (in $\mathrm{m}$).
The ionic strength $I$ of the solution is defined as
\begin{equation}
  I = \frac{1}{2} \left( \sum_{i=1}^N z_i^2 c_i + w \right).
\end{equation}

Finally, the Poisson equation for the unknown electric potential $\varphi$ reads
\begin{equation}
  - \underbrace{\ddiv(\epsilon \nabla \varphi)}_{\text{potential}}
  - \underbrace{F \left( \sum_{i=1}^N z_i c_i + w\right)}_{\text{electroneutrality}}
  = 0.
  \label{eq:pnp-poisson}
\end{equation}

In this example we consider a system with four ionic species: \ce{SO4^{2-}}, \ce{Mg^{2+}},
\ce{Na^{+}}, and \ce{K^{+}}. Their valences are $z_i = (-2, 2, 1, 1)$, their ionic radii are
$a_i = (6, 8, 4, 3) l_\text{ref}$ where $l_\text{ref} = \SI{1}{\angstrom}$ is the reference length
scale and their diffusion coefficients are $D_i = (3, 3, 3, 5) D_\text{ref}$ where
$D_\text{ref} = \SI{1e-10}{\meter^2/\second}$ is the reference diffusion coefficient. The ordering
in the vector notation corresponds to the species order above, i.e.
$i = (\ce{SO4^{2-}}, \ce{Mg^{2+}}, \ce{Na^{+}}, \ce{K^{+}})$. Importantly, we do not introduce each individual diffusion
coefficient and ionic radius as separate dimensional quantities, but we use reference quantities
$D_\text{ref}$ and $l_\text{ref}$ to express them in a nondimensional form. This choice is
justified because the diffusion coefficients and ionic radii differ only by small factors of order
$\mathcal O(1)$. In addition, this simplifies the dimensional analysis and reduces the number of
resulting $\Pi$ groups. Moreover, we look for a steady-state solution and simplify the mass conservation to
\begin{equation}
  \underbrace{\ddiv (D_i \nabla c_i)}_{\text{diffusion}}
  + \underbrace{\ddiv \left(\frac{z_i F}{R T} D_i c_i \nabla \varphi\right)}_{\text{convection}}
  + \underbrace{\ddiv (D_i c_i \nabla \ln(\gamma_i))}_{\text{Debye}} = 0, \quad i = 1, \ldots, 4.
\end{equation}

For brevity, we omit the definition of the dimensional quantities, which is similar
to the previous examples. The result of the automated Buckingham Pi dimensional analysis is shown in
Fig.~\ref{fig:pnp-pi-groups}.

\begin{figure}[h]
  \centering
  \begin{subfigure}[b]{\textwidth}
    \centering
    \begin{tabular}{@{}lll@{}}
      \toprule
      \textbf{Symbol}      & \textbf{Expression}                & \textbf{Value (base units)}                                             \\
      \midrule
      $c_\text{ref}$       & \SI{50}{\mole/\meter\cubed}        & \SI{50}{\mole/\meter\cubed}                                             \\
      $\varphi_\text{ref}$ & \SI{1}{\volt}                      & \SI{1}{\kilogram\meter\squared/(\ampere\second\cubed)}                  \\
      $D_\text{ref}$       & \SI{1e-10}{\meter\squared/\second} & \SI{1e-10}{\meter\squared/\second}                                      \\
      $\epsilon_0$         & \SI{8.854e-12}{\farad/\meter}      & \SI{8.854e-12}{\ampere\squared\second\tothe{4}/(\kilogram\meter\cubed)} \\
      $F$                  & \SI{9.649e4}{\coulomb/\mole}       & \SI{9.649e4}{\ampere\second/\mole}                                      \\
      $R$                  & \SI{8.314}{\joule/(\kelvin\mole)}  & \SI{8.314}{\kilogram\meter\squared/(\kelvin\mole\second\squared)}       \\
      $T$                  & \SI{300}{\kelvin}                  & \SI{300}{\kelvin}                                                       \\
      $l_\text{ref}$       & \SI{1}{\angstrom}                  & \SI{1e-10}{\meter}                                                      \\
      $e_0$                & \SI{1.602e-19}{\coulomb}           & \SI{1.602e-19}{\ampere\second}                                          \\
      \bottomrule
    \end{tabular}
    \subcaption{Physical quantities}
  \end{subfigure}

  \vspace{1em}

  \begin{subfigure}[b]{\textwidth}
    \centering
    \begin{tabular}{@{}lccccccccc@{}}
      \toprule
      \textbf{Dim.} & $c_\text{ref}$ & $\varphi_\text{ref}$ & $D_\text{ref}$ & $\epsilon_0$ & $F$  & $R$  & $T$ & $l_\text{ref}$ & $e_0$ \\
      \midrule
      N             & 1              & 0                    & 0              & 0            & $-1$ & $-1$ & 0   & 0              & 0     \\
      I             & 0              & $-1$                 & 0              & 2            & 1    & 0    & 0   & 0              & 1     \\
      L             & $-3$           & 2                    & 2              & $-3$         & 0    & 2    & 0   & 1              & 0     \\
      M             & 0              & 1                    & 0              & $-1$         & 0    & 1    & 0   & 0              & 0     \\
      $\Theta$      & 0              & 0                    & 0              & 0            & 0    & $-1$ & 1   & 0              & 0     \\
      T             & 0              & $-3$                 & $-1$           & 4            & 1    & $-2$ & 0   & 0              & 1     \\
      \bottomrule
    \end{tabular}
    \subcaption{Dimension matrix (only non-zero rows shown: J all zero)}
  \end{subfigure}

  \vspace{1em}

  \begin{subfigure}[b]{\textwidth}
    \centering
    \begin{tabular}{@{}lll@{}}
      \toprule
      \textbf{Group} & \textbf{Expression}                                                       & \textbf{Value} \\
      \midrule
      $\Pi_1$        & $R T/(F \varphi_\text{ref})$                                              & 0.0259         \\
      $\Pi_2$        & $\sqrt{F c_\text{ref}} l_\text{ref}/\sqrt{\epsilon_0 \varphi_\text{ref}}$ & 0.0738         \\
      $\Pi_3$        & $\sqrt{F c_\text{ref}} e_0/(\epsilon_0 \varphi_\text{ref})^{3/2}$         & 13.4           \\
      \bottomrule
    \end{tabular}
    \subcaption{Dimensionless groups}
  \end{subfigure}

  \caption{Buckingham Pi dimensional analysis for the Poisson--Nernst--Planck system. This output was generated by the automated dimensional analysis framework presented in this work.}
  \label{fig:pnp-pi-groups}
\end{figure}

The automated analysis identifies three dimensionless groups:
\begin{equation}
  \begin{aligned}
    \Pi_1 & = \frac{R T / F}{\varphi_\mathrm{ref}}                                                                                                                                            & \quad \text{(thermal voltage to reference voltage)}, \\
    \Pi_2 & = \frac{l_\text{ref}}{\sqrt{\epsilon_0 \varphi_\text{ref} / (F c_\text{ref})}}                                                                                                    & \quad \text{(reference length to Debye length)},     \\
    \Pi_3 & = \frac{e_0 \sqrt{F c_\text{ref}}}{(\epsilon_0 \varphi_\text{ref})^{3/2}} = \frac{e_0 / (\epsilon_0 \varphi_\text{ref})}{\sqrt{\epsilon_0 \varphi_\text{ref} / (F c_\text{ref})}} & \quad \text{(Coulomb length to Debye length)}.
  \end{aligned}
\end{equation}

The first group $\Pi_1$ represents the ratio of the thermal voltage $V_T = R T / F$ to the reference
voltage $\varphi_\mathrm{ref}$. This
number characterizes the energy scale of thermal fluctuations relative to the electric potential
energy set by the reference voltage.

The second group $\Pi_2$ compares the reference length scale to
the Debye length $\lambda_D = \sqrt{\epsilon_0 \varphi_\text{ref} / (F c_\text{ref})}$, which
indicates the distance over which electric potential decays in the electrolyte. A small value of
$\Pi_2$ suggests that the reference length is much smaller than the Debye length, implying weak
screening effects.

The third group $\Pi_3$ relates the Coulomb length to the Debye length. The
Coulomb length indicates the distance at which electrostatic interactions between ions become
significant compared to thermal energy. Note, that the standard definitions of the Debye length and
Coulomb length typically involve the thermal voltage $V_T$, but here they are expressed in terms of
the reference voltage $\varphi_\mathrm{ref}$. The relation to the standard definition can be recovered
with the help of the first Pi group $\Pi_1$. This mixing of the reference voltage and thermal
voltage arises as the consequence of the non-unique choice of the basis of the kernel of the dimension
matrix, as discussed multiple times in this work.

The Poisson--Nernst--Planck multiphysics example requires transformation, factorization and
normalization executed on two coupled variational forms independently: one for the Nernst-Planck
equations \eqref{eq:pnp-nernst-planck} and one for the Poisson equation \eqref{eq:pnp-poisson}.
These two weak forms have different physical dimensions and they require different reference factors
$\alpha_\mathrm{ref}$ for normalization.

Transformation step is based on the following mapping:
\begin{equation}
  \begin{aligned}
    c_i \mapsto c_\mathrm{ref} \, c_i, \quad
    \varphi \mapsto \varphi_\mathrm{ref} \, \varphi, \quad
    \delta c_i \mapsto c_\mathrm{ref} \, \delta c_i, \quad
    \delta \varphi \mapsto \varphi_\mathrm{ref} \, \delta \varphi,
  \end{aligned}
\end{equation}
which is again complemented by the spatial scaling of the differential operators and integration measures
with the reference length $l_\mathrm{ref}$. An important detail is the definition of the fixed charge concentration
$w$ appearing in both the Nernst--Planck and Poisson equations. We transform the fixed charge concentration
as $w \mapsto c_\mathrm{ref} \sum_{i=1}^N c^{(0)}_i z_i$, i.e. in terms of the reference concentration
$c_\mathrm{ref}$ and the initial concentrations $c^{(0)}_i$ of each species and their valences $z_i$.
This step is necessary to ensure that the electroneutrality condition and the ionic strength
are homogeneous in the reference concentration $c_\mathrm{ref}$. If the fixed charge concentration
was defined using an independent reference scale, say $w_\mathrm{ref}$, the dimensional factorization would
raise an exception due to inconsistent factors. Similarly, we express the ion radii $a_i$ using the
reference length $l_\mathrm{ref}$.

The Poisson equation weak form contains two terms labeled ``potential'' and ``electroneutrality'',
as denoted in Eq. \eqref{eq:pnp-poisson}. We choose the reference factor from the "potential" term,
which leads to the outcome as shown in Fig.~\ref{fig:pnp-poisson-terms}. The dimensionless group specific
to the Poisson equation is the square of the second Pi group $\Pi_2^2$, which appears in the
normalization of the ``electroneutrality'' term. This reflects the balance between the electric field
contribution and the charge density contribution in the Poisson equation.

\begin{figure}[h]
  \centering
  \begin{subfigure}[b]{\textwidth}
    \centering
    \begin{tabular}{@{}ll@{}}
      \toprule
      \textbf{Expression}         & $\epsilon_0 \varphi_\text{ref}^2 / l_\text{ref}$ \\
      \textbf{Value (base units)} & \SI{0.08854}{\kilogram/\second\squared}          \\
      \bottomrule
    \end{tabular}
    \subcaption{Reference factor from ``potential''}
  \end{subfigure}

  \vspace{0.5em}

  \begin{subfigure}[b]{\textwidth}
    \centering
    \begin{tabular}{@{}lll@{}}
      \toprule
      \textbf{Term}     & \textbf{Factor}                                                   & \textbf{Value} \\
      \midrule
      potential         & $1$                                                               & 1              \\
      electroneutrality & $F c_\text{ref} l_\text{ref}^2 / (\epsilon_0 \varphi_\text{ref})$ & 0.005449       \\
      \bottomrule
    \end{tabular}
    \subcaption{Normalized terms}
  \end{subfigure}

  \caption{Automated dimensional factorization and normalization for the Poisson equation weak
    form in the Poisson--Nernst--Planck system. Each term is expressed with respect to the reference
    factor from the potential term, resulting in dimensionless coefficients. This output was generated by the dimensional analysis
    framework presented in this work.}
  \label{fig:pnp-poisson-terms}
\end{figure}

The Nernst--Planck equations are less straightforward to factorize, due to the presence of the Debye
term $\nabla \ln(\gamma_i)$. Even though the activity coefficient $\gamma_i$ is dimensionless, and
so is its logarithm, the computation of $\ln(\gamma_i)$ from the extended Debye--Hückel model (Eq.~\ref{eq:pnp-debye-huckel})
involves the term $1 + B a_i \sqrt{I}$ in the denominator, which is not homogeneous in the chosen set of dimensional
quantities. The factorization process detects this inconsistency and raises an exception, similar to
the Neo-Hooke hyperelasticity case. To resolve this, we follow the similar approach to the Neo-Hooke example
and expand the extended Debye--Hückel model using a Taylor series for small ion radii $a_i$:
\begin{equation}
  \ln(\gamma_i) = -A z_i^2 \sqrt{I} + A B a_i z_i^2 I + \mathcal{O}(a_i^2).
\end{equation}
This expansion allows us to split the Debye term into two contributions: constant in
the ion radius $a_i$ and linear in $a_i$, which we term ``debye\_0th'' and ``debye\_1st'', respectively.

The result of the dimensional factorization and normalization for the Nernst--Planck weak form after choosing
the reference factor from the diffusion term is shown in Fig.~\ref{fig:pnp-nernst-planck-terms}.
\begin{figure}[h]
  \centering
  \begin{subfigure}[b]{\textwidth}
    \centering
    \begin{tabular}{@{}ll@{}}
      \toprule
      \textbf{Expression}         & $D_\text{ref} c_\text{ref}^2 / l_\text{ref}$        \\
      \textbf{Value (base units)} & \SI{2500.0}{\mole\squared/(\meter\tothe{5}\second)} \\
      \bottomrule
    \end{tabular}
    \subcaption{Reference factor from ``diffusion''}
  \end{subfigure}

  \vspace{0.5em}

  \begin{subfigure}[b]{\textwidth}
    \centering
    \begin{tabular}{@{}lll@{}}
      \toprule
      \textbf{Term} & \textbf{Factor}                                          & \textbf{Value} \\
      \midrule
      diffusion     & $1$                                                      & 1              \\
      convection    & $F \varphi_\text{ref} / (R T)$                           & 38.68          \\
      debye\_0th    & $F^2 \sqrt{c_\text{ref}} e_0 / (R T \epsilon_0)^{3/2}$   & 3213           \\
      debye\_1st    & $F^3 c_\text{ref} e_0 l_\text{ref} / (R T \epsilon_0)^2$ & 1475           \\
      \bottomrule
    \end{tabular}
    \subcaption{Normalized terms}
  \end{subfigure}

  \caption{Automated dimensional factorization and normalization for the Nernst--Planck equation weak
    form in the Poisson--Nernst--Planck system. Each term is expressed with respect to the reference
    factor from the diffusion term, resulting in dimensionless coefficients. This output was generated by the dimensional analysis
    framework presented in this work.}
  \label{fig:pnp-nernst-planck-terms}
\end{figure}
The dimensionless groups appearing in Fig.~\ref{fig:pnp-nernst-planck-terms} include the first group $\Pi_1$
in the convection term, reflecting the ratio of thermal voltage to reference voltage. The Debye terms
involve more complex combinations of the dimensional quantities, indicating their sensitivity to the
ionic strength and ion size effects. From the results in Figs. \ref{fig:pnp-poisson-terms} and
\ref{fig:pnp-nernst-planck-terms}, we observe that the most dominant contribution in the Nernst-Planck
weak form arises from the zeroth-order Debye term, followed by the first-order Debye term. We also note that
with the Taylor expansion, the Debye terms remain significantly larger than the diffusion and convection terms,
suggesting that ion-ion interactions play a crucial role in the transport dynamics within this example system.
Moreover, the first-order Debye term is still large compared to the diffusion term, indicating that even more
terms in the expansion may be necessary for an accurate representation of the activity coefficient effects.

We include the final dimensionless form of the coupled Poisson--Nernst--Planck system in Appendix
\ref{ap:pnp-dimless}. Let us emphasize that with the dimensional analysis framework
developed here, the final dimensionless form is automatically derived and hidden from the user.

%%%%%%%%%%%%%%%%%%%%%%%%%%%%%%%%%%%%%%%%%%%%%%%%%%%%%%%%%%%%%%%%%%%%%%%%%%%%%%%
%%%%%%%%%%%%%%%%%%%%%%%%%%%%%%%%%%%%%%%%%%%%%%%%%%%%%%%%%%%%%%%%%%%%%%%%%%%%%%%

\section{Conclusion}

This work establishes a rigorous framework for integrating dimensional analysis into the Unified
Form Language (UFL). We achieve this through symbolic quantity tracking and graph-based homogeneous
factorization.

The implementation leverages the interoperability between UFL and SymPy. We introduced a
\texttt{Quantity} class to associate physical units with UFL constants. We exploit the algebraic
structure of physical dimensions. Since they form an abelian group, we represent them as vectors in
$\mathbb{Q}^7$ within the SI system. This representation simplifies the implementation and enhances
performance. The core functionality relies on the visitor pattern to traverse the UFL expression
graph (DAG). This approach allows us to inspect and modify the computational graph directly.
Consequently, the software injects scaling factors and performs factorization automatically, without
requiring changes to the user's high-level mathematical syntax.

We demonstrated that automated nondimensionalization acts as a physics-aware diagonal
preconditioner. This process algebraically equilibrates the resulting linear systems. Crucially,
this method constitutes Full Operator Preconditioning (FOP). Scaling factors are applied directly to
the variational form prior to assembly. This preserves floating-point significance that is otherwise
lost in standard algebraic preconditioning.

Analysis of the incompressible Navier--Stokes equations confirmed the role of reference scales. We
showed that the condition number of the saddle-point matrix is explicitly controlled by
dimensionless groups, such as the Euler number. Furthermore, we applied the framework to the
Neo-Hooke hyperelasticity model. It successfully identified factor inconsistency in small
deformation gradients ($u_\text{ref} / l_\text{ref} \ll 1$) as an a priori indicator of catastrophic cancellation.
Consequently, the \texttt{dolfiny.units} module provides a necessary mechanism to enforce
dimensional consistency and optimize numerical stability at the formulation level.
In the Poisson--Nernst--Planck multiphysics system, we demonstrated the framework's capability to handle
complex coupled equations. The automated dimensional analysis identified dimensionless groups
that characterize the interplay between thermal, electric, and concentration effects.

The current implementation has a few limitations. Handling of vector and tensor quantities with associated
units is limited. Tensor-valued nodes are allowed, but they must have consistent units across all
components. Future work could extend support for tensors with different units in different
components. Additionally, not all UFL operators are natively supported. Nevertheless, it is easy for users to
extend the implementation by adding new visitor methods for unsupported operators.

Overall, we believe that the concepts presented in this work are useful regardless of the specific
implementation in UFL. The ability to automatically check for dimensional consistency, identify
dimensionless groups, and nondimensionalize variational forms is a powerful tool for scientific
computing. It helps to prevent modeling errors, improve numerical stability, and gain insights into
the physical system under investigation. We hope that similar capabilities will be integrated into
other scientific computing frameworks in the future.

\clearpage

\printbibliography

\appendix
\section{Estimate of the condition number for the Navier--Stokes system}
\label{ap:ns-condition-number}

Following is a standard result from the theory of saddle-point systems,
see \citet[\S 3.5]{Benzi2005Saddle}. First, we show that the smallest singular value
of the matrix $\mathbf K$ defined in Eq.~\eqref{eq:ns-saddle-point-system} approaches zero,
$\sigma_1 \to 0$, as $\mathrm{Eu} \to 0$. Consider
\begin{equation}
  \abs{\det \mathbf K} = \abs{\det \mathbf A \det(-\mathrm{Eu}^2 \mathbf B \mathbf A^{-1} \mathbf B^\trp)} =
  \mathrm{Eu}^{2m} \abs{\det \mathbf A \det(\mathbf B \mathbf A^{-1} \mathbf B^\trp)} = \mathrm{Eu}^{2m} C_0,
\end{equation}
which follows from the determinant expansion of the Schur factorization of $\mathbf K$, i.e.
\begin{equation}
  \mathbf K =
  \begin{pmatrix}
    \mathbf I                            & 0         \\
    \mathrm{Eu} \mathbf B \mathbf A^{-1} & \mathbf I
  \end{pmatrix}
  \begin{pmatrix}
    \mathbf A & 0                                                      \\
    0         & -\mathrm{Eu}^2 \mathbf B \mathbf A^{-1} \mathbf B^\trp
  \end{pmatrix}
  \begin{pmatrix}
    \mathbf I & \mathrm{Eu} \mathbf A^{-1} \mathbf B^\trp \\
    0         & \mathbf I
  \end{pmatrix}
\end{equation}
and $C_0$ is a constant independent of $\mathrm{Eu}$.
The largest singular value is bounded below. We have $\sigma_N \ge \norm{\mathbf A}$.
We use the ordering of singular
values $\sigma_1 \leq \sigma_2 \leq \ldots \leq \sigma_N$, thus
\begin{equation}
  \sigma_1^N \leq \sigma_1 \sigma_2 \ldots \sigma_N = \abs{\det \mathbf K}
\end{equation}
and consequently
\begin{equation}
  \sigma_1 \leq \abs{\det \mathbf K}^{1/N} = \mathrm{Eu}^{2m/N} C_0^{1/N}.
\end{equation}
Combining the upper bound for $\sigma_1$ with the lower bound for $\sigma_N$ leads to
\begin{equation}
  \kappa_2(\mathbf K) = \frac{\sigma_{N}}{\sigma_1} \geq
  \frac{\norm{\mathbf A}}{C_0^{1/N} \mathrm{Eu}^{2m/N}} = C_1 \mathrm{Eu}^{-2m/N},
\end{equation}
for some constant $C_1 > 0$ independent of $\mathrm{Eu}$.

\section{Expansion of Neo-Hooke strain energy density}
\label{ap:neo-hooke-expansion}

To resolve the factorization inconsistency without imposing artificial constraints on the reference
scales, we abandon the formulation based directly on the deformation gradient $\mathbf F$ and switch
to the Green-Lagrange strain tensor $\mathbf E$. While $\mathbf F$ describes the deformation
relative to the current configuration (scaling as $\mathbf I + \mathcal{O}(\Pi_3)$), the
Green-Lagrange strain operates on the displacement deviation directly, allowing the identity term to
be eliminated from the kinematic description.

The Green-Lagrange strain is defined as
\begin{align}
  \mathbf E = \frac{1}{2} (\mathbf C - \mathbf I) = \frac{1}{2} (\mathbf F^T \mathbf F - \mathbf I).
\end{align}
Substituting the definition $\mathbf F = \mathbf I + \nabla \mathbf u$, the strain tensor naturally
splits into linear and nonlinear contributions
\begin{equation}
  \begin{aligned}
    \mathbf E & = & \underbrace{\frac{1}{2}(\nabla \mathbf u + \nabla \mathbf u^T)}_{\text{linear part}}
              & + & \underbrace{\frac{1}{2}\nabla \mathbf u^T \nabla \mathbf u}_{\text{nonlinear part}}                     \\
              & = & \mathbf E_1                                                                          & + & \mathbf E_2.
  \end{aligned}
\end{equation}
Consequently, the strain energy density $\tilde W$ must be reformulated as a function of $\mathbf E$
rather than $\mathbf C$. We utilize the identities relating the invariants of $\mathbf C$ to
$\mathbf E$
\begin{align}
  I_1(\mathbf C) & = \mathrm{tr}(\mathbf I + 2\mathbf E) = 3 + 2\mathrm{tr}(\mathbf E), \\
  J(\mathbf C)   & = \sqrt{\det(\mathbf I + 2\mathbf E)}.
\end{align}
Substituting these into the original Neo-Hookean energy density expression yields
\begin{equation}
  \begin{aligned}
    \tilde W(\mathbf E) & = \frac{\mu}{2} \left( (3 + 2\mathrm{tr}(\mathbf E)) - 3 - 2 \ln(J(\mathbf E)) \right)
    + \frac{\kappa}{2} \left( J(\mathbf E) - 1 \right)^2                                                                                     \\
                        & = \underbrace{\mu \left( \mathrm{tr}(\mathbf E) - \ln(J(\mathbf E)) \right)}_{\text{isochoric term } \tilde W_\mu}
    + \underbrace{\frac{\kappa}{2} \left( J(\mathbf E) - 1 \right)^2}_{\text{volumetric term } \tilde W_\kappa}.
  \end{aligned}
\end{equation}

Under transformation $\mathbf u \mapsto u_\mathrm{ref} \mathbf u$ and $\nabla \mapsto (1/l_\mathrm{ref}) \nabla$,
the Green-Lagrange strain transforms as
\begin{align}
  \mathbf E  = \mathbf E_1 + \mathbf E_2 \mapsto \Pi_3 \mathbf E_1 + \Pi_3^2 \mathbf E_2.
\end{align}
We expand the strain energy density $\tilde W(\Pi_3 \mathbf E_1 + \Pi_3^2 \mathbf E_2)$ in terms of $\Pi_3$ around
$\Pi_3 = 0$. First, we use the determinant expansion
\begin{equation}
  \det(\mathbf I + 2 \mathbf E) = 1 + 2 \tr(\mathbf E) + 2 \tr(\mathbf E)^2 - 2 \tr(\mathbf E^2) + \mathcal O(\norm{\mathbf E}^3).
\end{equation}
Since the square root $\sqrt{1+x} = 1 + x/2 - x^2/8 + \mathcal{O}(x^3)$, we have
\begin{equation}
  J(\mathbf E) = 1 + \underbrace{\left(\tr(\mathbf E) + \tr(\mathbf E)^2 - \tr(\mathbf E^2)\right)}_{x/2}
  - \underbrace{\frac{1}{2}\tr(\mathbf E)^2}_{x^2/8 \approx (2 \tr(\mathbf E))^2/8} + \mathcal{O}(\norm{\mathbf E}^3),
\end{equation}
where we have only kept the term $2 \tr(\mathbf E)$ in the $x^2/8$ contribution, since the other terms are of higher order
in $\norm{\mathbf E}$. This can be regrouped as
\begin{equation}
  J(\mathbf E) = 1 + \tr(\mathbf E) + \frac{1}{2} \tr(\mathbf E)^2 - \tr(\mathbf E^2) + \mathcal{O}(\norm{\mathbf E}^3).
\end{equation}
This expansion of $J$ is accurate to second order in $\norm{\mathbf E}$, which is sufficient for the volumetric
term $(J-1)^2$ below. The isochoric term, however, requires $\ln(J)$ to third order, which cannot be recovered
from the second-order expansion of $J$ above. We therefore evaluate the logarithm directly through the matrix
identity $\ln(\det(\mathbf I + 2\mathbf E)) = \tr(\ln(\mathbf I + 2\mathbf E))$ together with the series
$\ln(\mathbf I + \mathbf A) = \mathbf A - \tfrac{1}{2}\mathbf A^2 + \tfrac{1}{3}\mathbf A^3 - \mathcal O(\norm{\mathbf A}^4)$.
With $\mathbf A = 2\mathbf E$ this gives
\begin{equation}
  \begin{aligned}
    \ln(J(\mathbf E)) & = \frac{1}{2} \ln(\det(\mathbf I + 2\mathbf E))
    = \frac{1}{2} \tr\left( 2\mathbf E - \frac{1}{2}(2\mathbf E)^2 + \frac{1}{3}(2\mathbf E)^3 \right) + \mathcal{O}(\norm{\mathbf E}^4) \\
                      & = \tr(\mathbf E) - \tr(\mathbf E^2) + \frac{4}{3} \tr(\mathbf E^3) + \mathcal{O}(\norm{\mathbf E}^4),
  \end{aligned}
\end{equation}
which we can insert into the isochoric part of the strain energy density
\begin{equation}
  \tilde W_\mu(\mathbf E) = \mu \left( \tr(\mathbf E) - \ln(J(\mathbf E)) \right)
  = \mu \left( \tr(\mathbf E^2) - \frac{4}{3} \tr(\mathbf E^3) \right) + \mathcal{O}(\norm{\mathbf E}^4).
\end{equation}
Using the expansion of the Green-Lagrange strain in terms of $\Pi_3$, we have
\begin{equation}
  \tilde W_\mu = \mu \left( \Pi_3^2 \tr(\mathbf E_1^2)
  + \Pi_3^3 \left( 2 \tr(\mathbf E_1 \mathbf E_2) - \frac{4}{3} \tr(\mathbf E_1^3) \right) \right)
  + \mathcal{O}(\Pi_3^4).
\end{equation}

The volumetric part $\tilde W_\kappa$ uses
\begin{equation}
  (J(\mathbf E) - 1)^2 = \tr(\mathbf E)^2 + \tr(\mathbf E)^3 - 2 \tr(\mathbf E) \tr(\mathbf E^2) + \mathcal{O}(\norm{\mathbf E}^4),
\end{equation}
which gives
\begin{equation}
  \tilde W_\kappa = \frac{\kappa}{2} \left( \Pi_3^2 \tr(\mathbf E_1)^2
  + \Pi_3^3 \left( 2 \tr(\mathbf E_1) \tr(\mathbf E_2) + \tr(\mathbf E_1)^3 - 2 \tr(\mathbf E_1) \tr(\mathbf E_1^2) \right)\right).
\end{equation}
Overall, the Neo-Hookean strain energy density expanded up to third order in $\Pi_3$ reads
\begin{equation}
  \begin{aligned}
    \tilde W       & = \tilde W^{(2)} + \tilde W^{(3)} + \mathcal{O}(\Pi_3^4),                                                                                             \\
    \tilde W^{(2)} & = \Pi_3^2 \bigl( \underbrace{\mu \tr(\mathbf E_1^2) + \frac{\kappa}{2} \tr(\mathbf E_1)^2}_{\tfrac12 \mathbf S_\text{lin} : \mathbf E_1} \bigr),      \\
    \tilde W^{(3)} & = \Pi_3^3 \bigl( \underbrace{\left( 2 \mu \mathbf E_1 + \kappa \tr(\mathbf E_1) \mathbf I \right) : \mathbf E_2}_{\mathbf S_\text{lin} : \mathbf E_2}
    + \kappa \left(\frac{1}{2}\tr(\mathbf E_1)^3 - \tr(\mathbf E_1) \tr(\mathbf E_1^2)\right) - \frac{4 \mu}{3} \tr(\mathbf E_1^3) \bigr).
  \end{aligned}
\end{equation}

\section{The dimensionless form of the Neo-Hooke hyperelasticity}
\label{ap:neo-hooke-dimless}

Dimensionless form of the compressible Neo-Hooke hyperelasticity based on the expansion around
$\Pi_3 = u_\text{ref} / l_\text{ref} = 0$ developed in
Appendix~\ref{ap:neo-hooke-expansion} and the automated nondimensionalization framework developed in
this work is given by: find $\mathbf u \in U_0$ which solves
\begin{equation}
  \begin{aligned}
    \min_{\mathbf u \in U_0} W(\mathbf u) - W_\text{ext}(\mathbf u)
  \end{aligned}
\end{equation}
where the dimensionless internal strain energy is taken as
\begin{equation}
  \begin{aligned}
    W(\mathbf u) = \frac{\mu}{\kappa}                                                                                        & \int_\Omega \tr(\mathbf E_1^2) \, \mathrm dx                                                                                                     & \text{(shear 2nd order)} \\
    +              \frac{1}{2}                                                                                               & \int_\Omega \tr(\mathbf E_1)^2 \, \mathrm dx                                                                                                     & \text{(bulk 2nd order)}  \\
    +                                 \frac{\mu}{\kappa} \frac{u_\text{ref}}{l_\text{ref}}                                   & \int_\Omega \left( 2 \mathbf E_1 : \mathbf E_2 - \frac{4}{3} \tr(\mathbf E_1^3) \right) \, \mathrm dx                                            & \text{(shear 3rd order)} \\
    +                                                                                      \frac{u_\text{ref}}{l_\text{ref}} & \int_\Omega \left(\tr(\mathbf E_1) \tr(\mathbf E_2) + \frac{1}{2} \tr(\mathbf E_1)^3 - \tr(\mathbf E_1) \tr(\mathbf E_1^2) \right) \, \mathrm dx & \text{(bulk 3rd order)}
  \end{aligned}
\end{equation}
and the dimensionless external work for a traction force $\mathbf t$ applied on the Neumann boundary $\Gamma_N$ is
\begin{equation}
  W_\text{ext}(\mathbf u) = \frac{l_\text{ref}}{u_\text{ref}} \frac{\tau_\text{ref}}{\kappa} \int_{\Gamma_N} \mathbf t \cdot \mathbf u \, \mathrm ds.
\end{equation}

In the above formulation, the only dimensional quantities are the shear modulus $\mu$, the bulk modulus
$\kappa$, the reference displacement $u_\text{ref}$, the reference length $l_\text{ref}$, and the reference
traction $\tau_\text{ref}$. All other quantities are dimensionless. There are three dimensionless
groups present in the formulation: the ratio of shear to bulk modulus $\mu / \kappa$, the ratio of
reference displacement to reference length $u_\text{ref} / l_\text{ref}$, and the ratio of reference
traction to bulk modulus $\tau_\text{ref} / \kappa$.

\section{The dimensionless form of the Poisson--Nernst--Planck equations}
\label{ap:pnp-dimless}

Dimensionless form of the steady-state Poisson--Nernst--Planck equations based on the automated
nondimensionalization framework developed in this work is given by: for $N$ species, find their
concentrations $c_i \in W_0$, $i=1,\ldots,N$ and the electric potential $\varphi \in V_0$ which solve
the coupled Poisson equation
\begin{equation}
  \begin{aligned}
                                                                          & \int_\Omega \epsilon_r \nabla \varphi \cdot \nabla \delta \varphi \, \mathrm dx & \text{ (potential) }         \\
    - \frac{F c_\text{ref} l_\text{ref}^2}{\epsilon_0 \varphi_\text{ref}} & \int_\Omega \left( \sum_{i=1}^N z_i c_i + w\right) \delta \varphi \, \mathrm dx & \text{ (electroneutrality) } \\
                                                                          & = 0 \quad \forall \delta \varphi \in V_0,
  \end{aligned}
\end{equation}
and the steady-state Nernst--Planck equations for each species $i=1,\ldots,N$
\begin{equation}
  \begin{aligned}
                                                                   & \int_\Omega \nabla c_i \cdot \nabla \delta c_i \, \mathrm dx                & \text{ (diffusion) }       \\
    + \frac{F \varphi_\text{ref}}{R T}                             & \int_\Omega z_i c_i \nabla \varphi \cdot \nabla \delta c_i \, \mathrm dx    & \text{ (convection) }      \\
    + \frac{F^2 e_0 \sqrt{c_\text{ref}}}{(R T \epsilon_0)^{3/2}}   & \int_\Omega c_i \nabla \beta_i^{(0)} \cdot \nabla \delta c_i \, \mathrm dx  & \text{ (0th order Debye) } \\
    + \frac{F^3 e_0 l_\text{ref} c_\text{ref}}{(R T \epsilon_0)^2} & \int_\Omega c_i \nabla \beta_i^{(1)} \cdot \nabla \delta c_i  \, \mathrm dx & \text{(1st order Debye) }  \\
                                                                   & = 0 \quad \forall \delta c_i \in W_0,
  \end{aligned}
\end{equation}
where we use the symbols $\beta_i^{(0)}$ and $\beta_i^{(1)}$ to denote the 0th and 1st order
Debye terms from the Taylor expansion of the activity coefficient $\ln(\gamma_i)$ defined as
\begin{equation}
  \begin{aligned}
    \beta_i^{(0)} & = - \frac{\sqrt{2}}{8 \pi \epsilon_r^{3/2}} z_i^2 \sqrt{I},                               \\
    \beta_i^{(1)} & = \frac{\sqrt{2}}{8 \pi \epsilon_r^{3/2}} \frac{\sqrt{2}}{\sqrt{\epsilon_r}} a_i z_i^2 I.
  \end{aligned}
\end{equation}

In the above formulation, the only dimensional quantities are the reference concentration
$c_\text{ref}$, the reference potential $\varphi_\text{ref}$, the reference length $l_\text{ref}$,
the absolute temperature $T$, the ionic radii $a_i$, the Faraday constant $F$, the universal gas
constant $R$, the vacuum permittivity $\epsilon_0$, and the elementary charge $e_0$. All other
quantities are dimensionless. The reference diffusion coefficient $D_\text{ref}$ is not present in
the dimensionless formulation, as it cancels out in the steady-state equations.

\end{document}